\documentclass[aps,pre,reprint,amsmath,amssymb,graphicx,longbibliography,superscriptaddress]{revtex4-2}

\usepackage{mathrsfs}
\usepackage[scr=rsfs,cal=boondox]{mathalfa}

\usepackage{bm}
\usepackage{amsmath}
\usepackage{amsfonts}
\usepackage[normalem]{ulem}
\usepackage[colorlinks=true,allcolors=blue,bookmarks=false,pdfusetitle]{hyperref}
\usepackage{url}
\usepackage{xcolor} 
\usepackage{graphicx}
\usepackage{breqn}
\usepackage{braket}
\usepackage{cases}
\usepackage[utf8]{inputenc}
\usepackage[T1]{fontenc}

\usepackage[overload]{empheq}
\usepackage{cleveref}

\usepackage{tikz}

\makeatletter
\let\cat@comma@active\@empty
\makeatother

\begin{document}

\title{Enhanced performance of sudden-quench quantum Otto cycles via multi-parameter control}

\author{R. S. Watson}
\affiliation{School of Mathematics and Physics, University of Queensland, Brisbane,  Queensland 4072, Australia}
\author{K. V. Kheruntsyan}
\affiliation{School of Mathematics and Physics, University of Queensland, Brisbane, Queensland 4072, Australia}

\date{\today{}}

\begin{abstract}
\noindent Advances in experimental control of interacting quantum many-body systems with multiple tunable parameters—such as ultracold atomic gases and trapped ions—are driving rapid progress in quantum thermodynamics and enabling the design of quantum thermal machines.
In this work, we utilize a sudden quench approximation as a means to investigate the operation of a quantum thermodynamic Otto cycle in which multiple parameters are simultaneously controllable. The method applies universally to many-body systems where such control is available, and therefore provides general principles for investigating their operation as a working medium in quantum thermal machines. 
We investigate application of this multi-parameter quench protocol in an experimentally realistic one-dimensional Bose gas, as well as in the transverse-field Ising model. We find that such a multi-parameter Otto cycle, when operating as an engine, outperforms not only its constituent single-parameter Otto cycles in terms of the net work and efficiency, but also the combined net work of its constituent engine cycles when added together independently. We also find that a similar multi-parameter enhancement applies to the coefficient of performance when the Otto cycle operates as a refrigerator.
\end{abstract}

\maketitle

\section{Introduction}\label{sec:intro}
Out-of-equilibrium dynamics of interacting many-body systems are at the forefront of both theoretical and experimental quantum physics, spurred in large part by rapid advancement in the experimental control over quantum platforms \cite{Rossnagel2016single,Ca-ion-spin-engine,Maslennikov2019,VanHorne2020,Nitrogen-vacancy-heat-engine,bouton2021quantum,horvath2025observingbethestringsattractive}. Investigation into such systems offers unprecedented opportunities within the field of quantum thermodynamics, which itself is a rapidly developing field with an emphasis on understanding thermodynamic principles arising within a quantum context \cite{vinjanampathy2016quantum,kosfloff2014quantum,potts2024quantum,Bhattacharjee2021,Gluza2021,deffner2019quantum,campbell2025roadmapquantumthermodynamics}. Indeed, the realization of, and precise control over, many-body interacting quantum systems represents an important step in the study and understanding of thermodynamics in complex quantum systems.

Engine cycles have been central to the historical development of quantum thermodynamics, with origins dating back to 1953 in Scovil and Schulz-Dubois' analysis of the maser as a single-body 3-level quantum heat engine \cite{scovil2959three}. Recently, in order to advance the understanding of quantum engines and their scaling, many researchers have focused on the operation of quantum devices in quantum many-body systems \cite{koch2022making,simmons2023thermodynamic,bouton2021quantum,jaramillo2016quantum,watson2025quantum,Nautiyal_2024,watson2025universal,le2018spin,williamson2023manybody,sajitha2025quantum,halpern2019quantum,chen2019interaction,keller2020feshbach,atas2020thermo,brollo2025universalefficiencyboostprethermal}. Such many-body interacting quantum systems enable investigation into the role of uniquely quantum effects in engine operation, such as quantum coherence \cite{narasimhachar2015low,Korzekwa2016extraction,
lostaglio2015description,bernardo2020unraveling,kammerlander2016coherence,Cwiklinski2015limitations} or correlations \cite{oppenheim2002thermo,llobet2015extractable,Huber_2015,Williamson_2025,watson2025quantum}.

Notable recent examples have been the experimental realizations of quantum Otto engines in interacting ultracold Bose gases \cite{koch2022making,simmons2023thermodynamic}. These experiments exploited quasistatic control over both interparticle interaction strength and the frequency of an external harmonic trap in an alternating fashion, thus realizing \textit{uniquely} quantum many-body thermodynamic devices. In particular, Ref.~\cite{koch2022making} elegantly demonstrated the role that quantum statistics played in the performance of their quasistatic quantum Otto engine cycle. 

Extending the investigation of quantum many-body engine cycles to non-quasistatic protocols, i.e. to out-of-equilibrium regimes of operation, is essential in order to understand their performance in scenarios that generate a finite power output, which vanishes in the quasistatic limit. However, such an extension remains challenging due to the typical complexity of simulating the out-of-equilibrium dynamics of interacting many-body systems in experimentally realistic parameter regimes, where precise and rapid control is available over various system parameters \cite{smith2019simulating,vidal2004efficient,dufty2020generalized,doyon2025generalized,atas2020thermo,boubakour2023interaction,Nautiyal_2024}.
An extreme version of an out-of-equilibrium engine operation is realized via a \emph{sudden} quench, where one can approximate the final post-quench state as unchanged from its initial thermal equilibrium state in order to again rely on exact thermal equilibrium expectation values for calculation of the net work and efficiency \cite{watson2025quantum,watson2025universal}.

In this work, we examine a quantum Otto engine cycle, as well as other related thermal machines, such as an Otto refrigerator, operating under a sudden quench of multiple externally tunable parameters. In particular, we extend the recent work on Otto engine cycles, operating under a \emph{single} parameter quench for arbitrary quantum models in Ref.~\cite{watson2025universal}, to the case of simultaneous quenching of \emph{multiple} control parameters. We apply this formalism to the harmonically trapped one-dimensional (1D) Bose gas and the transverse-field Ising model (TFIM), where we observe a general enhancement to the net work under two-parameter control when contrasted with the combined performance of two single-parameter quenches.

\section{Multi-parameter sudden quench Otto engine}
\label{sec:mutli_param}

We begin by describing a general physical model that may correspond to, e.g., an ultracold quantum gas or a spin chain on a lattice, which incorporates multiple externally controllable parameters. In detail, in the context of second quantized quantum mechanics, we define a model Hamiltonian that consists first of the operators $\{\hat{\mathcal{V}}^{(\alpha)}\}$ (indexed by $\alpha=1,2,3,...$) intended to correspond to all operators with external control over a scalar strength parameter, denoted $\{c^{(\alpha)}\}$, such that the Hamiltonian contains the terms $c^{(\alpha)}\hat{\mathcal{V}}^{(\alpha)}$.
The second set of operators, collectively denoted $\hat{H}_0$, corresponds to any and all terms without externally controlled strength parameters, a common example of which is the kinetic energy.
Thus, the full Hamiltonian is given by,
\begin{equation}\label{eq:hamiltonian_multiparameter}
    \hat{H} = \sum_{\alpha} c^{(\alpha)} \, \hat{\mathcal{V}}^{(\alpha)} + \hat{H}_0 .
\end{equation}
We further note that the set $\{\hat{\mathcal{V}}^{(\alpha)}\}$ may consist of any combination of one-body operators, two-body operators, etc., and likewise any product of spin operators for spin models describing quantum magnetism.

The sudden quench approximation utilized in this work consists of assuming that the period of time over which the set of parameters $\{c^{(\alpha)}\}$ is quenched between the initial ($i$) and final ($f$) values, e.g. $\{c^{(\alpha)}_i\}\!\to \!\{c^{(\alpha)}_f\}$, is significantly faster than the timescale over which the density matrix of the initial equilibrium state, $\hat{\rho}_i$, is capable of responding. 
Under such an approximation, the energy of the state immediately after the quench may be evaluated as
\begin{equation}
    \langle \hat{H}\rangle_f \simeq \langle \hat{H}_0 \rangle_i +  \sum_\alpha c^{(\alpha)}_f   \big\langle \hat{\mathcal{V}}^{(\alpha)} \big\rangle_i, 
\end{equation}
where $\langle \hat{\mathcal{O}}\rangle_i \!=\! \mathrm{Tr} [ \hat{\rho}_i\hat{\mathcal{O}}]$ denotes the expectation value of the operator $\hat{\mathcal{O}}$ in the equilibrium state defined by the initial density matrix $\hat{\rho}_i$.
The work of the corresponding unitary stroke in the Otto engine cycle (see below), $W_{i\to f} \!\equiv\!\langle\hat{H} \rangle_f \!-\! \langle\hat{H} \rangle_i$, may therefore be approximated as
\begin{equation}\label{eq:W_if}
    W_{i\to f}\simeq \sum_\alpha (c^{(\alpha)}_f - c^{(\alpha)}_i)  \big\langle \hat{\mathcal{V}}^{(\alpha)} \big\rangle_i,
\end{equation} 
where the expectation value of the uncontrolled terms is unchanged, $\langle \hat{H}_0\rangle_f\!=\!\langle \hat{H}_0\rangle_i$, and therefore cancels when taking the difference $\langle\hat{H} \rangle_f \!-\! \langle\hat{H} \rangle_i$ in calculating the work. The work $W_{i\to f}$ is therefore determined entirely from the knowledge of expectation values $\big\langle \hat{\mathcal{V}}^{(\alpha)} \big\rangle_i$ calculated for the initial equilibrium state $\hat{\rho}_i$. We note that this method was recently utilized in Refs.~\cite{watson2025quantum,watson2025universal} to evaluate the performance of an Otto engine cycle under a sudden quench of the interparticle interaction strength, in which case the net work was shown to be proportional to the same-position particle-particle  pair correlation function.

The multi-parameter quantum Otto engine cycle (described in greater detail in Appendix \ref{app:Otto_cycle}) operates between high ($h$) and low ($l$) energy equilibrium states, facilitated by coupling of the working fluid to two external reservoirs. Generally, the type of coupling can be either thermal or diffusive, or both. Such an engine cycle consists of four strokes: two unitary work strokes in which the working fluid is isolated from all external reservoirs and the externally controlled parameters are suddenly quenched between $\{ c_l^{(\alpha)}\}\rightleftarrows\{ c_h^{(\alpha)}\}$; and two equilibration strokes where the working fluid is coupled with one of the two external reservoirs and the externally controlled parameters, $\{c^{(\alpha)}\}$, are kept constant at their post-quench values.

The net work, $W$, achieved during this engine cycle is calculated from the work input and output generated during the course of the two unitary work strokes described above, i.e., $W \!=\! W_{\mathbf{A}\to \mathbf{B}} \!+\! W_{\mathbf{C}\to \mathbf{D}}$. As each unitary work stroke is evaluated using the sudden quench approximation, given by Eq.~\eqref{eq:W_if}, the total net work becomes:
\begin{equation}\label{eq:Work_generalised}
    W \simeq - \sum_{\alpha}(c_h^{(\alpha)} -c_l^{(\alpha)}) \left(\big\langle \hat{\mathcal{V}}^{(\alpha)} \big\rangle_h -  \big\langle \hat{\mathcal{V}}^{(\alpha)} \big\rangle_l \right).
\end{equation}
Such an Otto engine cycle generates net beneficial work (done by the fluid) when $W \!<\!0$. One may additionally evaluate the generalized engine efficiency, $\eta \!=\!-W/E_{\mathbf{B}\to\mathbf{C}}$ \cite{SchroederD_ThermalPhysics}, using the calculated energy intake $E_{\mathbf{B}\to\mathbf{C}}$ during the equilibration stroke with the high energy reservoir, which is equal to $E_{\mathbf{B}\to\mathbf{C}}\!=\!\langle \hat{H}\rangle_h-(\,\langle \hat{H}\rangle_l + W_{\mathbf{A} \to \mathbf{B}} \,)$. This generalized efficiency is then given by
\begin{equation}
\begin{split}
\begin{aligned}\label{eq:efficiency_generalised}
    \eta \simeq 1 - \frac{\langle\hat{H}\rangle_{\mathit{h}} - \langle\hat{H}\rangle_{\mathit{l}} - \sum_{\alpha} (c^{(\alpha)}_h - c^{(\alpha)}_l)\big\langle \hat{\mathcal{V}}^{(\alpha)} \big\rangle_h}{\langle\hat{H}\rangle_h - \langle\hat{H}\rangle_l - \sum_{\alpha} (c^{(\alpha)}_h - c^{(\alpha)}_l) \big\langle \hat{\mathcal{V}}^{(\alpha)} \big\rangle_l},
\end{aligned}
\end{split}
\end{equation}
where we used the conservation of energy $W\!+\!E \!=\! 0$, with $E = E_{\mathbf{B}\to\mathbf{C}}\! +\! E_{\mathbf{D}\to\mathbf{A}}$ being the total energy.

Importantly, the energy exchanged with the reservoirs during the equilibration strokes may take any form (e.g. heat, chemical work, etc.), requiring only that the set of strength parameters, $\{ c^{(\alpha)}\}$, are kept constant. The case where energy intake and output is given entirely by heat corresponds to a conventional quantum Otto heat engine cycle, in which case the generalized efficiency is equivalent to the standard thermodynamic efficiency \cite{SchroederD_ThermalPhysics,CallenHerbertB1985Taai}. Non-standard, or generalized, efficiency is utilized in cycles where the system-reservoir contact contains non-thermal components. A simple example of non-thermal contact is diffusive contact, which may be achieved through an exchange of particles between the system and reservoirs during the equilibration strokes; such a device is typically referred to as a \emph{thermo-chemical} Otto engine \cite{luo2015efficiency,marzolino2024thermo,Nautiyal_2024,watson2025universal}.

In the following we investigate the conditions under which this multi-parameter quantum Otto cycle exhibits an unambiguous benefit over single-parameter operation. Following this, in Secs.~\ref{sec:1D_bose_gas}-\ref{sec:finite_temp_1D_Bose}, we explore a particular case of this multi-parameter sudden quench Otto cycle in an experimentally realizable harmonically trapped 1D Bose gas. Such a system is amenable to rapid experimental control over both the frequency of the external harmonic trap and over the strength of interparticle interactions \cite{horvath2025observingbethestringsattractive,malvania2020generalized,haller2009realization,fang2014quench,chen2019interaction}. Finally, in Sec.~\ref{sec:TFIM}, we demonstrate application of multi-parameter Otto engine to the TFIM, additionally utilizing small parameter quenches, through which we explore the role of criticality for this two-parameter Otto cycle.

\section{Multi-parameter enhancement}\label{sec:general_multi_param}

For quantum systems where multiple strength parameters, $c^{(\alpha)}$, may be simultaneously suddenly quenched, one may generally consider the scenario where each parameter drives its own Otto cycle, while leaving all other parameters constant. From this perspective, it is interesting to consider whether operating the multi-parameter Otto cycle produces a benefit over the sum of these single-parameter cycles.

In order to examine the differences between the general multi-parameter Otto cycle and its single-parameter counterparts, we begin by defining the dependence of the equilibrium operator expectation values, $\langle \hat{\mathcal{V}}^{(\alpha)}\rangle$, on the strength parameters, $c^{(\alpha)}$. In particular, for a multi-parameter sudden quench Otto cycle where $M$ parameters are simultaneously quenched, it is useful to write the equilibrium expectation value of each operator explicitly as a function of all parameters,
\begin{equation}
    \langle \hat{\mathcal{V}}^{(\alpha)}\rangle\left(c^{(1)},c^{(2)},\dots,c^{(M)};T\right) .
\end{equation}
Doing so highlights the fact that each expectation values depend on \emph{all} parameters defining the equilibrium state, in addition to the temperature, $T$.

We may then define the expectation value of each operator within the Hamiltonian in the low energy equilibrium state,
\begin{equation}
    \langle \hat{\mathcal{V}}^{(\alpha)}\rangle_l \equiv  \langle \hat{\mathcal{V}}^{(\alpha)}\rangle\left(c^{(1)}_l,\dots,c^{(M)}_l;T_l\right),
\end{equation}
where all controllable parameters and the temperature are taken at their low energy values, $c^{(\alpha)}_l$ and $T_l$.
The high energy equilibrium state, $h$, may be described similarly,
\begin{equation}
    \langle \hat{\mathcal{V}}^{(\alpha)}\rangle_h \equiv  \langle \hat{\mathcal{V}}^{(\alpha)}\rangle\left(c^{(1)}_h,\dots,c^{(M)}_h;T_h\right).
    \label{eq:V_h}
\end{equation}

Using these notations, we express the net work of the \emph{sub-cycle} associated only with the single $c^{(\alpha)} \, \hat{\mathcal{V}}^{(\alpha)}$ term in the Hamiltonian,
\begin{equation}\label{eq:W_multi_alpha}
    W^{(\alpha)}_{\mathrm{multi}} = \left(c^{(\alpha)}_h - c^{(\alpha)}_l \right) \left(\langle \hat{\mathcal{V}}^{(\alpha)}\rangle_h - \langle \hat{\mathcal{V}}^{(\alpha)}\rangle_l \right).
\end{equation}
The total net work of the multi-parameter quench, with all $c^{(\alpha)}$'s being quenched between their respective $l$ to $h$ values, is additive, and hence
\begin{equation}
    W_{\mathrm{multi}} = \sum_{\alpha=1}^M W^{(\alpha)}_{\mathrm{multi}}.
\end{equation}

In contrast, one may define the single-parameter Otto cycle, generated via control over \emph{only} a single strength parameter, $c^{(\alpha)}$, while leaving all others constant at their low energy values. An example of such an Otto cycle is the volumetric Otto cycle in an interacting 1D Bose gas, where only the frequency of the external harmonic trap is controlled, and the inter-particle interaction strength left constant. Otto cycles of this type were previously investigated in Ref.~\cite{watson2025universal} for arbitrary quantum models under a sudden quench approximation.

In order to express the net work of the single-parameter Otto cycle associated with a quench of the control parameter $c^{(\alpha)}$ from $c^{(\alpha)}_l$ to $c^{(\alpha)}_h$, we define the high-energy equilibrium expectation value  for the  corresponding operator,
\begin{equation}
    \langle \hat{\mathcal{V}}^{(\alpha)}\rangle_{h(\alpha)} \!\equiv \! \langle \hat{\mathcal{V}}^{(\alpha)}\rangle\left(c^{(1)}_l,\dots,c^{(\alpha)}_h,\dots,c^{(M)}_l;T_h\right) .
\end{equation}
Importantly, all parameters \emph{except} $c^{(\alpha)}$ are kept constant at their low energy ($l$) values. The fact that only $c^{(\alpha)}$ is being quenched is reflected by the use of subscript $h(\alpha)$ in $\langle \hat{\mathcal{V}}^{(\alpha)}\rangle_{h(\alpha)} $, instead of simply $h$ as in Eq.~\eqref{eq:V_h}.
The net work of this single-parameter quench Otto cycle, denoted by $W_{\mathrm{single}}^{(\alpha)}$, is then given by
\begin{equation}\label{eq:W_single_alpha}
    W^{(\alpha)}_{\mathrm{single}} = \left(c^{(\alpha)}_h - c^{(\alpha)}_l \right) \left(\langle \hat{\mathcal{V}}^{(\alpha)}\rangle_{h(\alpha)} - \langle \hat{\mathcal{V}}^{(\alpha)}\rangle_{l}\right).
\end{equation}
Compared to Eq.~\eqref{eq:W_multi_alpha}, the superscript $(\alpha)$ here refers to the quenched parameter $c^{(\alpha)}$ (from the $l$ to $h$ value, while all the others remain at their $l$ values), and the net work comes again from a single $c^{(\alpha)} \, \hat{\mathcal{V}}^{(\alpha)}$ term in the Hamiltonian. We define the \emph{total} net work due to quenching all parameters $c^{(\alpha)}$ individually as $W_{\mathrm{single}}$,
\begin{equation}
    W_{\mathrm{single}}=\sum_{\alpha=1}^M W^{(\alpha)}_{\mathrm{single}}.
\end{equation}

We may now compare the net work of the multi-parameter Otto cycle with the sum of the net work achieved by the single-parameter cycles by their difference,
\begin{align}
    -\Delta W \equiv -W_{\mathrm{multi}} - \left(-W_{\mathrm{single}}   \right),
\end{align}
We define a multi-parameter enhancement to the net work by the condition $-\Delta W \!>\!0$. We then substitute the expressions for the net work of both the multi- and single-parameter Otto cycles, given in Eq.~\eqref{eq:W_multi_alpha} and Eq.~\eqref{eq:W_single_alpha}, respectively, to find 
\begin{align}\label{eq:Delta_W_out}
    -\Delta W \!=\! \sum_{\alpha=1}^M \left(c^{(\alpha)}_h \!-\! c^{(\alpha)}_l\right) \left(\big\langle \hat{\mathcal{V}}^{(\alpha)} \big\rangle_{h} \!-\! \big\langle \hat{\mathcal{V}}^{(\alpha)} \big\rangle_{h(\alpha)} \right).
\end{align}
As $c^{(\alpha)}_h\!>\!c^{(\alpha)}_l$ for all $\alpha$ by definition, the first bracketed term is always positive. Hence, a sufficient condition for multi-parameter enhancement of the net work is that $\big\langle \hat{\mathcal{V}}^{(\alpha)} \big\rangle_{h} \!>\! \big\langle \hat{\mathcal{V}}^{(\alpha)} \big\rangle_{h(\alpha)}$.

Although the net work can be expressed as a sum of contributions associated with each control parameter, the corresponding expectation values are not independent. Each expectation value depends on all control parameters through the many-body state (i.e., the many-body density operator used for evaluating the expectation values). Consequently, varying one parameter modifies the expectation values entering the work contribution of another parameter. The enhancement therefore arises from this mutual dependence of thermodynamic observables on multiple control parameters, rather than from inherently quantum resources such as coherence. In this sense, the apparent deviation from simple additivity reflects the fact that the relevant degrees of freedom are not independent.

We note that the multi-parameter enhancement mechanism discussed here does not rely on uniquely quantum features. Rather, it originates from the dependence of thermodynamic observables on multiple control parameters, such that changing one parameter modifies the expectation values associated with another. This mechanism is therefore, in principle, applicable also to classical interacting systems. Quantum many-body systems, however, provide a natural setting where such cross-dependencies are strong and experimentally accessible.

\subsection{Small quench sizes}\label{sec:small_quench}

When studying quantum Otto cycles, it is often useful to examine the case of small quench sizes, as such scenarios are commonly tractable to analytic methods. 
We therefore begin by defining the infinitesimal quench of the parameters $c^{(\alpha)}$ as $ c^{(\alpha)}_h\!-\!c^{(\alpha)}_l\!\equiv\!d c^{(\alpha)}$.
As noted previously, to have a multi-parameter enhancement to the net work, $-\Delta W\!>\!0$, one must be able to demonstrate that the second bracketed term given in Eq.~\eqref{eq:Delta_W_out} is positive. For small quenches, we may re-express this difference as,
\begin{equation}
    \big\langle \hat{\mathcal{V}}^{(\alpha)} \big\rangle_{h} - \big\langle \hat{\mathcal{V}}^{(\alpha)} \big\rangle_{h(\alpha)} \simeq \sum_{\beta \neq \alpha} d c^{(\beta)} \frac{\partial \langle \hat{\mathcal{V}}^{(\alpha)} \big\rangle_{h}}{\partial c^{(\beta)}},
\end{equation}
where we have utilized the fact that the difference between these terms arises from the change in all parameters \emph{except} $c^{(\alpha)}$.
Substituting this expression into our formula for net work enhancement, given in Eq.~\eqref{eq:Delta_W_out}, we find the approximate expression to second order,
\begin{equation}
    -\Delta W \simeq \sum_\alpha d c^{(\alpha)}\sum_{\beta \neq \alpha} d c^{(\beta)} \frac{\partial \langle \hat{\mathcal{V}}^{(\alpha)}\rangle_h}{\partial c^{(\beta)}}.
\end{equation}

In the thermodynamic limit, the Hellmann-Feynman theorem allows for expression of the expectation value in a thermal equilibrium state as \cite{hellmann1933,feynman1939forces,kheruntsyan2005finite}
\begin{equation}
    \langle \hat{\mathcal{V}}^{(\alpha)}\rangle_h = \frac{\partial F_h}{\partial c^{(\alpha)}},
\end{equation}
where $F_h$ is the Helmholtz free energy in the high energy equilibrium state. We may therefore re-express the net work enhancement for small parameter quenches as
\begin{equation}\label{eq:delta_W_smallquench}
     -\Delta W = 2\sum_{\alpha < \beta} d c^{(\alpha)} d c^{(\beta)} \frac{\partial^2 F_h}{\partial c^{(\alpha)}\partial c^{(\beta)}},
\end{equation}
where we have used the commutativity of partial derivatives to combine all pairs of mixed partial derivatives.

The second derivative of the free energy is directly related to the static susceptibility \cite{CallenHerbertB1985Taai,Taggart1979stability}.
As such, an interesting future direction would therefore be to extend the above to finite-time dynamics through linear response theory.
In this paper though, we will utilize these small parameter quench results in Sec.~\ref{sec:TFIM} for obtaining additional analytic insights for the TFIM, where the Helmholtz free energy at finite temperature, and hence its derivatives with respect to the Hamiltonian parameters, are know analytically.

\section{The 1D Bose gas}\label{sec:1D_bose_gas}
The Lieb-Liniger model of a 1D Bose gas is a paradigmatic quantum many-body model describing an ultracold atomic gas with repulsive contact pairwise interactions, being both experimentally realizable and having exact theoretical results \cite{liebliniger,yang1969thermodynamics,GHD_onatomchip,horvath2025observingbethestringsattractive,Kinoshita1125,Wilson1461,malvania2020generalized,GHD_onatomchip,Schmiedmayer_1D_PRL_2010,hofferberth2007non}. This model, in the uniform limit, consists first of an uncontrolled Hamiltonian operator, $\hat{H}_0$, corresponding to the kinetic energy,
\begin{equation}
\begin{split}
\begin{aligned}\label{eq:H_0_LL}
    \hat{H}_0 = -\frac{\hbar^2}{2m} \int dx \hat{\Psi}^\dagger(x) \frac{\partial^2}{\partial x^2} \hat{\Psi}(x).
\end{aligned}
\end{split} 
\end{equation} 
where $m$ is the atomic mass, and $\hat{\Psi}^{(\dagger)}(x)$ are the bosonic field annihilation (creation) operator at position $x$.
Contact ($\delta$-function) interactions are incorporated in Eq.~\eqref{eq:hamiltonian_multiparameter} via the interaction term
\begin{equation}
\label{eq:g2_LL}
    c^{(2)}\hat{\mathcal{V}}^{(2)}=\frac{g}{2}\hat{\overline{G_2}}=\frac{g}{2}\!\int dx \hat{G}_2(x),
\end{equation}
where the two-body operator $\hat{\mathcal{V}}^{(2)}\equiv\hat{\overline{G_2}}=\int dx \hat{G}_2(x)$ corresponds to the \emph{integrated} local (same-point) pair correlation $\hat{G}_2(x)\equiv\hat{G}_2(x,x)\!=\!\hat{\Psi}^{\dagger}(x)\hat{\Psi}^{\dagger}(x)\hat{\Psi}(x)\hat{\Psi}(x)$, which we note does not depend on $x$ in a uniform system due to the translational invariance.
The interaction strength $c^{(2)}$ (with $c^{(2)}>0$ for repulsive interactions) is given by  $c^{(2)}\!\equiv\!g/2\simeq \! \hbar \omega_\perp a_s$ away from confinement induced resonances \cite{Olshanii1:998} and is experimentally controllable via the tight transverse trapping frequency, $\omega_\perp$, or magnetic Feshbach resonance \cite{chin2010feshbach} tuning of the 3D $s$-wave scattering length $a_s$.
We may utilize the normalized local (same-point) two-body correlation function,
\begin{align}\label{eq:g2_function}
    g^{(2)}(x) &=  \frac{\langle \hat{G}_2(x) \rangle}{\rho(x)^2} =\frac{  \langle \hat{\Psi}^\dagger(x) \hat{\Psi}^\dagger(x) \hat{\Psi}(x) \hat{\Psi}(x) \rangle}{\rho(x)^2},
\end{align}
to express the expectation value of the integrated correlation function as $\langle \hat{\overline{G_2}}\rangle\!=\!\int dx \,g^{(2)}(x) \rho(x)^2 $, where $\rho(x)=\langle \hat{\rho}(x)\rangle=\langle \hat{\Psi}^\dagger(x) \hat{\Psi}(x) \rangle$ is the average particle number density. In the following, we utilize this to analytically evaluate the integrated local pair correlation function, taking advantage of known expressions for both the equilibrium density profile, $\rho(x)$ \cite{pethick2008bose,pitaevskii2016bose}, and for the normalized two-body correlation function, $g^{(2)}(x)$, in the relevant parameter regimes \cite{gangardt2003local,gangardt2003stability,kheruntsyan2003pair,kheruntsyan2005finite,kerr2023analytic}.

The uniform 1D Bose gas, with density $\rho\!=\!N/L$, at finite temperature is characterized by a dimensionless interaction strength, $\gamma\!=\!m g / \hbar^2 \rho$, and a dimensionless temperature, $\tau \!=\!T/T_d$, where $T_d \!=\!\hbar^2 \rho^2 / 2 m k_B $ is the temperature of quantum degeneracy. This model becomes analytically tractable in six distinct regimes as a function of $\gamma$ and $\tau$  \cite{kheruntsyan2003pair,kheruntsyan2005finite,kerr2023analytic}. 
More generally, it is numerically solvable at all interaction strengths and temperatures in the thermodynamic limit through the thermodynamic Bethe ansatz (TBA) \cite{yang1969thermodynamics}. We note here that the interaction-driven quantum Otto cycle for this model in its uniform limit has been previously investigated under both adiabatic \cite{chen2019interaction} and sudden quench \cite{watson2025quantum,watson2025universal} protocols.

Experimental realization of the 1D Bose gas typically occurs within an external harmonic trapping potential \cite{GHD_onatomchip,horvath2025observingbethestringsattractive,Schmiedmayer_1D_PRL_2010}, which is expressed via the one-body operator in Eq.~\eqref{eq:hamiltonian_multiparameter},
\begin{equation}
        c^{(1)}\hat{\mathcal{V}}^{(1)}\equiv \frac{1}{2}m\omega^2\int dx \,  x^2 \hat{\rho}(x),
\end{equation}
with $\hat{\mathcal{V}}^{(1)}\!\equiv\!\frac{1}{2}m\int dx \,  x^2 \hat{\rho}(x)$, where $c^{(1)}\!\equiv\! \omega^2 $ corresponds to the harmonic trapping frequency squared and $\hat{\rho}(x) \!=\!\hat{\Psi}^\dagger(x)\hat{\Psi}(x)$ is the particle number density operator. We additionally define the atomic position variance, $\langle x^2 \rangle \!\equiv \!\int dx \,  x^2 \langle \hat{\rho}(x) \rangle$, such that $\langle \hat{\mathcal{V}}^{(1)}\rangle \!=\!m\langle x^2\rangle/2$.

In the presence of an external trapping potential, one may again utilize the numerically exact TBA introduced above, along with the additional assumption of a local density approximation \cite{kheruntsyan2005finite}, to accurately model the non-uniform system at finite temperature. Through the external harmonic trap, the  density profile gains a dependence on the position, $\rho \!=\!\rho(x)$. The gas may then be described in terms of the dimensionless interaction strength and temperature at the trap center, i.e. $\gamma_0\!=\!mg/\hbar^2 \rho(0)$ and $\tau_0\!=\!2 m k_B T/\hbar^2 \rho(0)^2$, and the average total number of particles in the system, $N\!=\!\int dx \langle \hat{\rho}(x) \rangle$  \cite{kheruntsyan2005finite}. 
The interaction-driven and separately the volumetric (i.e. harmonic frequency quench) quantum Otto cycles for this system have been previously investigated under both quasistatic and sudden quench protocols \cite{Nautiyal_2024,watson2025universal,keller2020feshbach,simmons2023thermodynamic,koch2022making}.

Here, we instead consider a scenario with rapid control over \emph{both} the interaction strength and harmonic trapping frequency between two fixed values, denoted $(g_l,g_h)$ and $(\omega^2_{l},\omega^2_{h})$, with $g_l\!\leq\!g_h$ and $\omega^2_l\!\leq\!\omega^2_h$. This  enables realization of the two-parameter sudden quench quantum Otto engine cycle. The net work of such an engine cycle is calculated via Eq.~\eqref{eq:Work_generalised} (see Appendix \ref{app:Otto_cycle}), giving
\begin{align}\label{eq:work_bosegas}
    -W \simeq &\frac{1}{2} (g_h - g_l)\left( \langle \hat{\overline{G_2}}\rangle_h - \langle \hat{\overline{G_2}}\rangle_l \right)\nonumber \\ &+\frac{1}{2} m (\omega_h^2 - \omega_l^2)\left( \langle x^2\rangle_h - \langle x^2 \rangle_l \right).
\end{align}
Though the cycles are inherently interconnected, with the thermal equilibrium expectation values $\langle \hat{\overline{G_2}}\rangle_{h(l)}$ and $\langle x^2\rangle_{h(l)}$ depending on both controllable parameters $g$ and $\omega$, we separate this formula into an interaction-driven sub-cycle, $-W^g \!=\!\frac{1}{2} (g_h - g_l)\big( \langle \hat{\overline{G_2}}\rangle_h - \langle \hat{\overline{G_2}}\rangle_l \big)$, and a volumetric sub-cycle, $-W^\omega=\frac{1}{2} m (\omega_h^2 - \omega_l^2)\left( \langle x^2\rangle_h - \langle x^2 \rangle_l \right)$, such that $-W\!\simeq\!-(W^g + W^\omega)$. Such a partition will be useful later for examining the enhancement of engine performance when considering a simultaneous quench of both controllable parameters relative to single-parameter quenches.

The net work can therefore be deduced either in the ground state system, where the temperatures of the external reservoirs, and therefore the working fluid, are fixed to $T=0$, and only particle exchange takes place between the working fluid and the reservoirs, or at finite temperatures via TBA calculation, where both heat and particles may be exchanged. In both cases, one may calculate the equilibrium expectation values of both $\langle \hat{\overline{G_2}}\rangle$ and $\langle x^2 \rangle$ in the high- and low-energy equilibrium states. 
Further, there is a corresponding result for the generalized efficiency, given by Eq.~\eqref{eq:efficiency_generalised}, again calculable via the TBA.
We note that interaction enhancement of a volumetric Otto engine cycle was previously studied in the context of an adiabatic quantum Otto cycle for few-body interacting systems in Ref.~\cite{boubakour2023interaction}.

\subsection{Two-parameter enhancement}\label{sec:two_param}

Before turning to explicit calculation of the net work and efficiency, we provide a demonstration of the general multi-parameter enhancement to the net work for this two-parameter quantum Otto cycle.
For clarity, we emphasize the dependence of the equilibrium states $(h)$ and $(l)$ on the interaction strength and harmonic trapping frequency as $\langle \cdot \rangle_{(g,\omega)}$.
Using this notation, the formula for multi-parameter enhancement of the net work, $-\Delta W$, given in Eq.~\eqref{eq:Delta_W_out}, is expressed for the harmonically trapped 1D Bose gas as,
\begin{align}
    -\Delta W &=  \frac{1}{2}(g_h - g_l) \left(\langle \hat{\overline{G_2}}\rangle_{(g_h,\omega_h)} - \langle \hat{\overline{G_2}}\rangle_{(g_h,\omega_l)} \right) \nonumber \\
    &+  \frac{1}{2}(\omega_h^2 - \omega_l^2) \left(\langle x^2\rangle_{(g_h,\omega_h)} - \langle x^2\rangle_{(g_l,\omega_h)} \right).
\end{align}
This expression enables a direct comparison between the net work of the single-parameter Otto engines with the sub-cycles of the two-parameter Otto engine, $W^g$ and $W^\omega$, where $W\!=\!W^g\!+\!W^\omega$, as defined after Eq.~\eqref{eq:work_bosegas}.

To this end, we first note that any increase to the harmonic trapping frequency, $\omega_l\!\to\!\omega_h$ with $\omega_h\!>\!\omega_l$, for fixed interaction strength $g_h$, compresses the atomic cloud, thereby increasing the integral of the squared density profile, $\int dx \, \rho(x)^2$. To guarantee an overall increase to the integrated correlation function, given below Eq.~\eqref{eq:g2_function} as
$\langle \hat{\overline{G_2}}\rangle\!=\!\int dx \,g^{(2)}(x) \rho(x)^2 $, we additionally require that $g^{(2)}(x)$ grows under this same compression. Indeed, we know that $g^{(2)}(x)$ monotonically decreases with the dimensionless interaction strength $\gamma(x) \!=\!mg/\hbar^2 \rho(x)$ \cite{kerr2023analytic}. As this dimensionless interaction strength is inversely proportional to the density, we find that the total integrated correlation function indeed increases with the trapping frequency, meaning $\langle \hat{\overline{G_2}}\rangle_{(g_h,\omega_h)}\!>\! \langle \hat{\overline{G_2}}\rangle_{(g_h,\omega_l)}$ for $\omega_h\!>\!\omega_l$.

Similarly, increasing the interaction strength, $g_l\!\to\!g_h$ with $g_h\!>\!g_l$, at a fixed value of the harmonic trapping frequency, $\omega_h$, broadens the atomic cloud, as increased interparticle repulsion drives the atoms further apart. This results in an increased second moment of the density distribution, $\langle x^2\rangle$, meaning $\langle x^2\rangle_{(g_h,\omega_h)}\!>\!\langle x^2\rangle_{(g_l,\omega_h)}$.

Thus, for the case of a sudden quench of both interaction strength and harmonic trapping frequency in a 1D Bose gas, the net work of the two-parameter quantum Otto engines is greater than the sum of that for the single-parameter Otto engines taken in isolation. We additionally note that the same arguments apply to any Otto engine where control is over both interaction strength and volume of the gas. In particular, the arguments presented above also apply to the the 1D Bose gas confined to either a uniform box trap or within a ring, where the total system length is controllable. Such a scenario may be thought of as a quantum analogue of the classical piston engine.

We note that the enhanced performance of a two-parameter quench thermal machine over its single-parameter constituents, expressed as $-\Delta W\!>\!0$, remains true regardless of whether the Otto cycle operates as an engine (which is what we are considering here, in the main text), refrigerator, thermal accelerator, or heater \cite{watson2025quantum}. We illustrate this for the case of an Otto refrigerator in Appendix \ref{app:refrigerator}.

\section{Zero temperature quasicondensate regime}
To gain analytical insight and obtain transparent quantitative results on the enhancement of the Otto engine performance under a two-parameter quench, we first consider a simple example of a harmonically trapped, zero-temperature 1D Bose gas in the weakly interacting regime, with the dimensionless interaction strength in the trap centre satisfying $\gamma_0\ll1$) \cite{Petrov_2000_Regimes,Mora-Castin-2003,kheruntsyan2005finite}.

Since the working fluid is in its zero temperature ground state, the Otto cycle under consideration cannot operate as a heat engine (which cycles between cold and hot thermal equilibrium states via exchange of \emph{heat} with the cold and hot reservoirs). Instead, we examine its operation in the \emph{chemical} engine scenario \cite{chen1997chemical,hooyberghs2013chemical,luo2015efficiency,marzolino2024thermo,Nautiyal_2024}, facilitated by chemical work due to exchange of particles when the working fluid is in contact with the reservoirs for attaining the low- ($l$) and high-energy ($h$) equilibrium states.

At zero temperature, the weakly interacting Bose gas may is described by a Thomas-Fermi approximation \cite{pitaevskii2016bose,pethick2008bose}. From this, one is able to calculate both equilibrium expectation values, $\langle x^2\rangle$ and $\langle\hat{\overline{G_2}}\rangle$, present in Eq.~\eqref{eq:work_bosegas}  analytically. 
More specifically, for the expectation value $\langle x^2\rangle$ we obtain $\langle x^2\rangle \!=\!(N/5)R_{\mathrm{TF}}^2$, where $R_{\mathrm{TF}}\!=\!(3Ng/2m\omega^2)^{1/3}$ is the Thomas-Fermi radius \cite{pethick2008bose} (for further details, see Appendix \ref{app:thomas_fermi}).
For the correlation function $\langle\hat{\overline{G_2}}\rangle$, on the other hand, we obtain the approximation $\langle\hat{\overline{G_2}}\rangle    \simeq\!g^{(2)}(0)\int dx \rho(x)^2  \!\simeq\! b N  \rho(0) g^{(2)}(0)$, where $\rho(0)\!=\!(9mN^2\omega^2/32g)^{1/3}$ is the density of the Thomas-Fermi profile at the trap center, and $b$ is a constant factor determined by the form of the density profile, with $b\!=\!4/5$ for the Thomas-Fermi inverted parabola.  
Thus, the integrated correlation function depends on
\begin{align}
    g^{(2)}(0) &= \frac{  \langle \hat{\Psi}^\dagger(0) \hat{\Psi}^\dagger(0) \hat{\Psi}(0) \hat{\Psi}(0 \rangle}{\rho(0)^2},
\end{align}
which is the normalized local two-body correlation function at the trap center, i.e. at $x\!=\!0$.
In the weakly interacting ground state, we may approximate this correlation function as $g_{l(h)}^{(2)}(0)\!\simeq\! 1\!-\!2\sqrt{\gamma_{0,l(h)}}/\pi$ \cite{kerr2023analytic}, where $\gamma_{0,l(h)}\!=\!mg_{l(h)}/\hbar^2 \rho_{l(h)}(0)$ is the dimensionless interaction strength at the trap center in the low (high) energy equilibrium state, which depends on the peak density of the equilibrium density profile, $\rho_{l(h)}(x)$.

Finally, to also evaluate engine efficiency via Eq.~\eqref{eq:efficiency_generalised}, we require an expression for the total energy, $\langle \hat{H}\rangle_{l(h)}\!=\!\langle \hat{V}\rangle_{l(h)} + (g_{l(h)}/2)\langle \hat{\overline{G_2}}\rangle_{l(h)}$, in the low (high) energy equilibrium states. This can be expressed in the Thomas-Fermi approximation as
\begin{equation}\label{eq:energy_TF}
    \langle \hat{H}\rangle_{l(h)} \!=\! E^{\mathrm{TF}}_{l(h)} \!+\! \frac{g_{l(h)}}{2} b N_{l(h)} \rho_{l(h)}(0) \left(g^{(2)}_{l(h)}(0) \!-\! 1\right),
\end{equation}
where we have defined $E^{\mathrm{TF}}_{l(h)}\!=\!3 N_{l(h)} g_{l(h)} \rho_{l(h)}(0)/5$, as the total energy in the low (high) energy equilibrium state of the harmonically trapped 1D Bose gas that is fully coherent, i.e. $g^{(2)}_{l(h)}(0) \!=\!1$ \cite{pethick2008bose,pitaevskii2016bose}. The second term contains corrections to the interaction energy due to the approximate expression for the correlation function, where $g^{(2)}_{l(h)}(0) \!-\! 1\!\simeq\!-2\sqrt{\gamma_{0,l(h)}}/\pi$.

\begin{figure}[!t]
\centering
\includegraphics[width=0.47\textwidth]{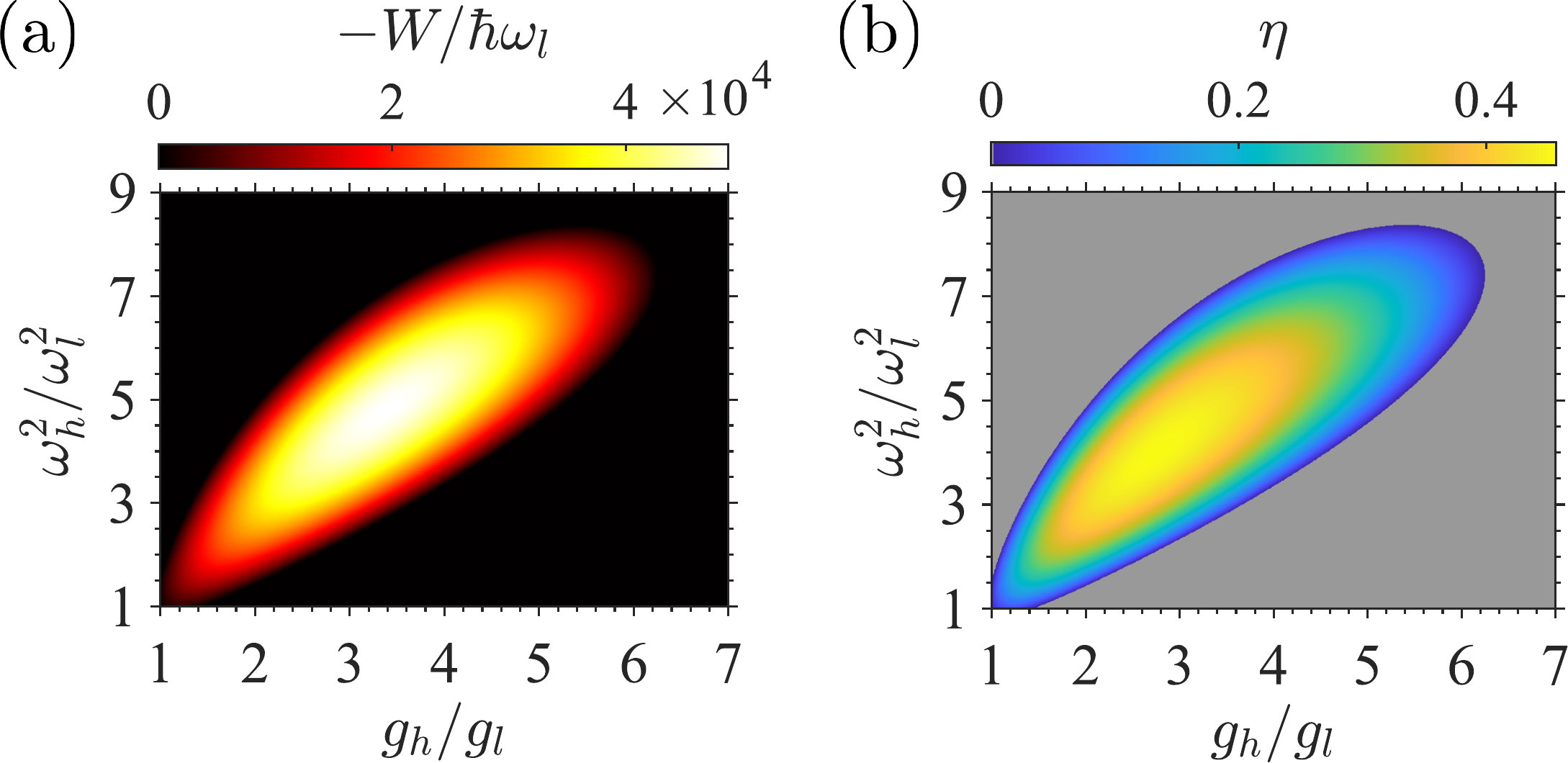}
\caption{Performance of the two-parameter sudden quench quantum Otto engine cycle in the weakly interacting ground state of the 1D Bose gas using the Thomas-Fermi approximation. 
Panels (a) and (b) demonstrate the net work ($W$) and efficiency ($\eta$), respectively, as a function of both the interaction strength ratio, $g_h/g_l$, along the horizontal axis, and harmonic trapping frequency ratio, $\omega_h^2/\omega_l^2$, along the vertical axis. The net work is represented in harmonic oscillator units defined by the longitudinal frequency in the low energy equilibrium state, $\omega_l$. The low energy equilibrium state ($l$) is parameterized by $N\!=\!2000$ total particles at dimensionless interaction strength $\gamma_0\!=\!4.9\times 10^{-2}$. $\Delta N\!=\!200$ particles are exchanged with the reservoirs while in contact.}
\label{fig:GP_T0}
\end{figure}

Combining these approximations in Eqs.~\eqref{eq:efficiency_generalised} and \eqref{eq:work_bosegas}, we thus obtain simple analytic results for the efficiency and net work of the two-parameter sudden quench chemical Otto engine and illustrate them in Fig.~\ref{fig:GP_T0}, as functions of both quenched parameter ratios, $g_h/g_l$ and $\omega_h^2/\omega_l^2$.
Here, the system is initialized in the low-energy equilibrium state with $N\!=\!2000$ particles. Then, following the work input stroke, $W_{\mathbf{A}\to\mathbf{B}}$, the working fluid takes in $\Delta N\!=\!200$ particles during the equilibration stroke $\mathbf{B}\!\to\!\mathbf{C}$ with the high energy reservoir. To operate in a closed cycle, the same number of particles is later output into the low energy reservoir during the corresponding equilibration stroke $\mathbf{D}\!\to\!\mathbf{A}$ \cite{particle_exchange}.

The net work of this two-parameter sudden quench Otto cycle, shown in Fig.~\ref{fig:GP_T0}(a), indicates that engine operation occurs only within a finite region of the quenched parameter ratios, with a maximum net work achieved at the center of this region. In particular, the maximum net work achieved is $-W/\hbar \omega_l \!\simeq\!4.9\times 10^{4} $, which, when normalized to the total particle number in the low energy equilibrium state, corresponds to $-W/N_l\hbar \omega_l \!\simeq\!25 $. Further, we find an efficiency at maximum net work of $\eta \!\simeq\!0.42$. Normalized to total particle number in the low energy equilibrium state,  Importantly, this performance represents a significant enhancement over the \emph{single-parameter} sudden quench quantum Otto engine cycles investigated recently for the 1D Bose gas in Ref.~\cite{watson2025universal}.

\begin{figure}[!t]
\centering
\includegraphics[width=0.47\textwidth]{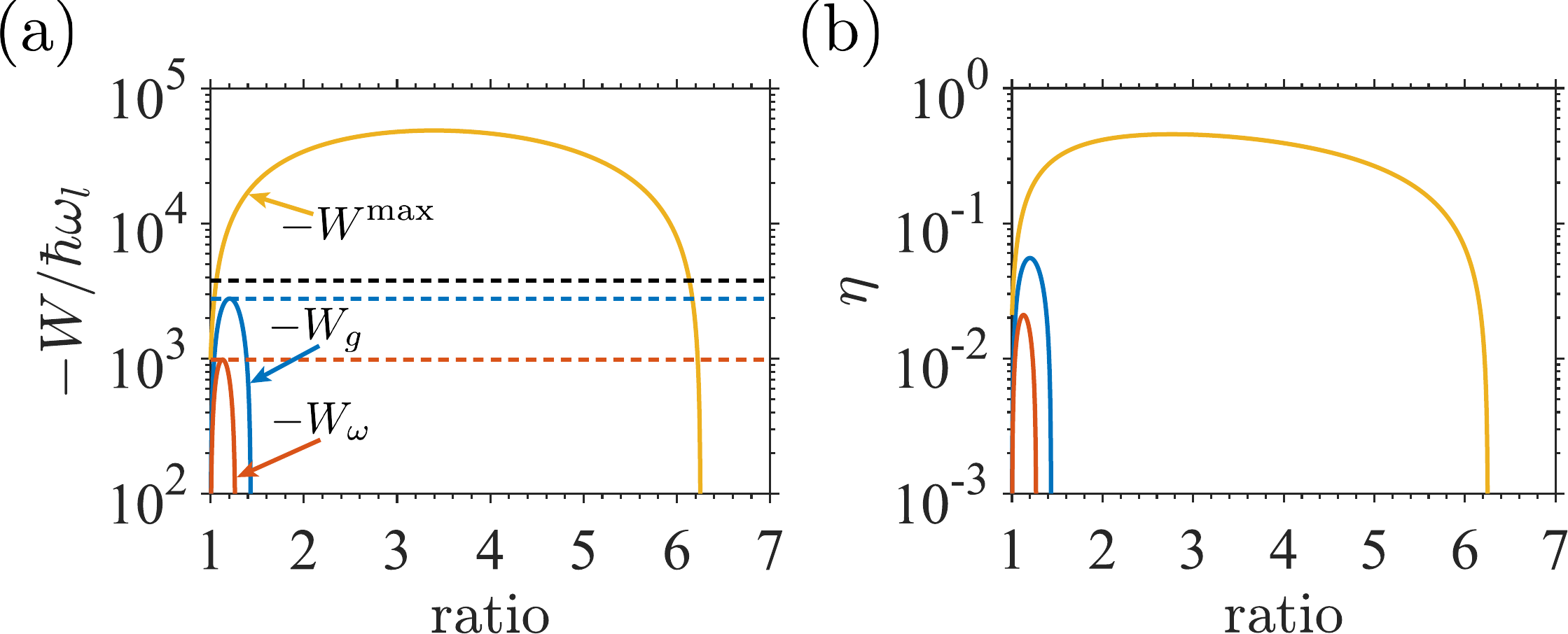}
\caption{Comparison of the two-parameter sudden quench quantum Otto engine cycle against the single-parameter Otto engine cycles. In panel (a), we demonstrate the net work of the single-parameter Otto engine cycles as a function of their quenched parameter; the interaction-driven cycle, $W_g$, is shown as the solid blue line, whereas the volumetric cycle, $W_\omega$, as the solid red line. The value of the quenched parameter ratio is denoted `ratio'. This is contrasted with the maximum net work of the two-parameter Otto engine cycle, denoted $W^{\mathrm{max}}$ and shown as the solid yellow line. The maximum net work of both single-parameter Otto engines are given by the blue and red dashed lines for the interaction-driven and volumetric cycles, respectively, with their sum shown as the black dashed line. In contrast, the maximum net work of the two-parameter Otto engine cycle is greater than this sum by more than an order of magnitude. Panel (b) demonstrates the efficiency of the single-parameter Otto engine cycles, with colors corresponding to those shown in panel (a). The efficiency at maximum net work of the two-parameter Otto engine is shown as the solid yellow line, and clearly out-performs both single-parameter cycles both in terms of magnitude and breadth of operation.}
\label{fig:GP_T0_compare}
\end{figure}

In particular, the net work and efficiency under control of only the ratio $g_h/g_l$ (or $\omega_h^2/\omega_l^2$) while keeping the complementary quantity, $\omega_h^2/\omega_l^2\!=\!1$ (or $g_h/g_l\!=\!1$), constant, are shown in Fig.~\ref{fig:GP_T0_compare} (a) and (b). The quenched parameter is denoted as the `ratio' on the horizontal axis of this figure, and denotes either $g_h/g_l$ for $-W_g$, or $\omega_h^2/\omega_l^2$ for $-W_\omega$. For comparison, we plot the maximum net work of the two-parameter Otto engine cycle, $-W^{\mathrm{max}}$, which is found by calculating the volumetric ratio $\omega_h^2/\omega_l^2$ that gives the maximum net work for each value of the interaction strength ratio $g_h/g_l$.

Previously, in Sec.~\ref{sec:two_param}, we have shown that, for a working fluid consisting of a harmonically trapped 1D Bose gas, the performance of the two-parameter Otto cycle generally out-performs the sum of the individual cycles taken in isolation. Here, in this concrete example of a two-parameter Otto engine, we see that this engine \emph{greatly} outperforms the individual Otto engines, as the two-parameter cycle generates net work that is more than an order of magnitude greater than the sum of both individual engine cycles operating at their respective maxima (shown as the black dashed line in Fig.~\ref{fig:GP_T0_compare}(a)). Additionally, though it was not guaranteed by the analysis presented in Sec.~\ref{sec:two_param}, we find that the efficiency of the two-parameter Otto engine cycle also outperforms what is achieved by both single-parameter engine cycles.

To explain why engine operation only occurs over a region of the parameter space, as shown in Fig.~\ref{fig:GP_T0} (a), we decompose the net work into two contributions,
\begin{align}\label{eq:W_coh_corr}
    W \simeq W_{\mathrm{coh.}} + W_{\mathrm{corr.}},
\end{align}
where $W_{\mathrm{coh.}}$ denotes the extractable net work from a fully coherent gas, i.e. approximating $g^{(2)}_l(0)\!=\!g^{(2)}_h(0)\!\simeq\! 1$. In particular, utilizing the analytic expressions derived from the Thomas-Fermi approximation, the net work extracted from a fully coherent working fluid is given by
\begin{align}
    W_{\mathrm{coh.}} =& -\frac{b}{2}(g_h-g_l) \left(N_h \rho_h(0) - N_l \rho_l(0) \right) \nonumber \\
    &-\frac{1}{10} m (\omega_h^2 - \omega_l^2) \left( N_h R_h^2 - N_l R_l^2 \right).
\end{align}
The second term in Eq.~\eqref{eq:W_coh_corr} contains higher order terms arising from corrections to the local second order correlation function due to the finite interaction strength,
\begin{align}\label{eq:W_corr}
    W_{\mathrm{corr.}} \!=\! \frac{b}{\pi} (g_h \!-\! g_l) \left( N_h \rho_h(0) \sqrt{\gamma_{0,h}} \!-\! N_l \rho_l(0) \sqrt{\gamma_{0,l}}\right).
\end{align}
We now analyze these two contributions in greater detail.

To simplify our analysis, we consider the net work as a function of equal parameter ratios, that is for $g_h/g_l\!=\!\omega_h^2/\omega_l^2 \!\equiv\! r$. Applying this to the coherent contribution to the net work, we find
\begin{align}
    W_{\mathrm{coh.}} &\!\simeq\!  -E^{\mathrm{TF}}_l  (r \!-\! 1) \left( \frac{N_h^{5/3}}{N_l^{5/3}} \!-\! 1\right).
\end{align}
Therefore, assuming that we are operating with a finite quench size, $r\!>\!0$, and with non-zero particle intake, $\Delta N=N_h\!-\!N_l>0$, the coherent contribution to the net work guarantees engine operation, $W\!<\!0$.

The additional terms contributing to the net work, given in Eq.~\eqref{eq:W_corr}, contain terms arising from the corrections to the coherent correlations, i.e. from  $g^{(2)}(0)\!-\!1\!=\!2\sqrt{\gamma(0)}/\pi$. It is these corrections to the correlation function that restrict engine operation to a limited range of the parameter ratios. In particular, the dimensionless interaction strength of the high energy equilibrium state, denoted $\gamma_{0,h}$, scales as $\gamma_{0,h}\!\propto \!r$, causing this contribution to reduce the overall net work. At large enough values of this ratio, these correction terms result in $W\!>\!0$, meaning the cycle no longer operates as an engine.
We therefore note that taking account of these corrections to coherent correlations is essential to correctly evaluating the performance of this quantum Otto engine.

\section{Finite temperature 1D Bose gas}\label{sec:finite_temp_1D_Bose}

\subsection{Quasicondensate}
At low but finite temperatures such that $2\gamma_0\!\ll\!\tau_0 \!\ll\! 2\sqrt{\gamma_0}$, the weakly interacting ($\gamma_0\!\ll\! 1$) 1D Bose gas inhabits the thermal quasicondensate regime \cite{kerr2023analytic}. Here, the Thomas-Fermi approximation for the density profile is no longer a good approximation, and must be replaced by the thermodynamic Bethe ansatz solution under a local density approximation \cite{yang1969thermodynamics,kheruntsyan2005finite}. This method is numerically exact, and enables calculation of both equilibrium expectation values required to evaluate engine performance via Eqs.~\eqref{eq:efficiency_generalised} and \eqref{eq:work_bosegas}. At non-zero temperature, we may additionally utilize a finite temperature difference between the two equilibrium states $\mathbf{A}$ and $\mathbf{C}$ (see Appendix \ref{app:Otto_cycle}), meaning our working fluid exchanges both heat and particles with the reservoirs during equilibration, i.e. we investigate operation of a \emph{thermo-chemical} Otto engine cycle \cite{Nautiyal_2024}.

\begin{figure}[!t]
\centering
\includegraphics[width=0.48\textwidth]{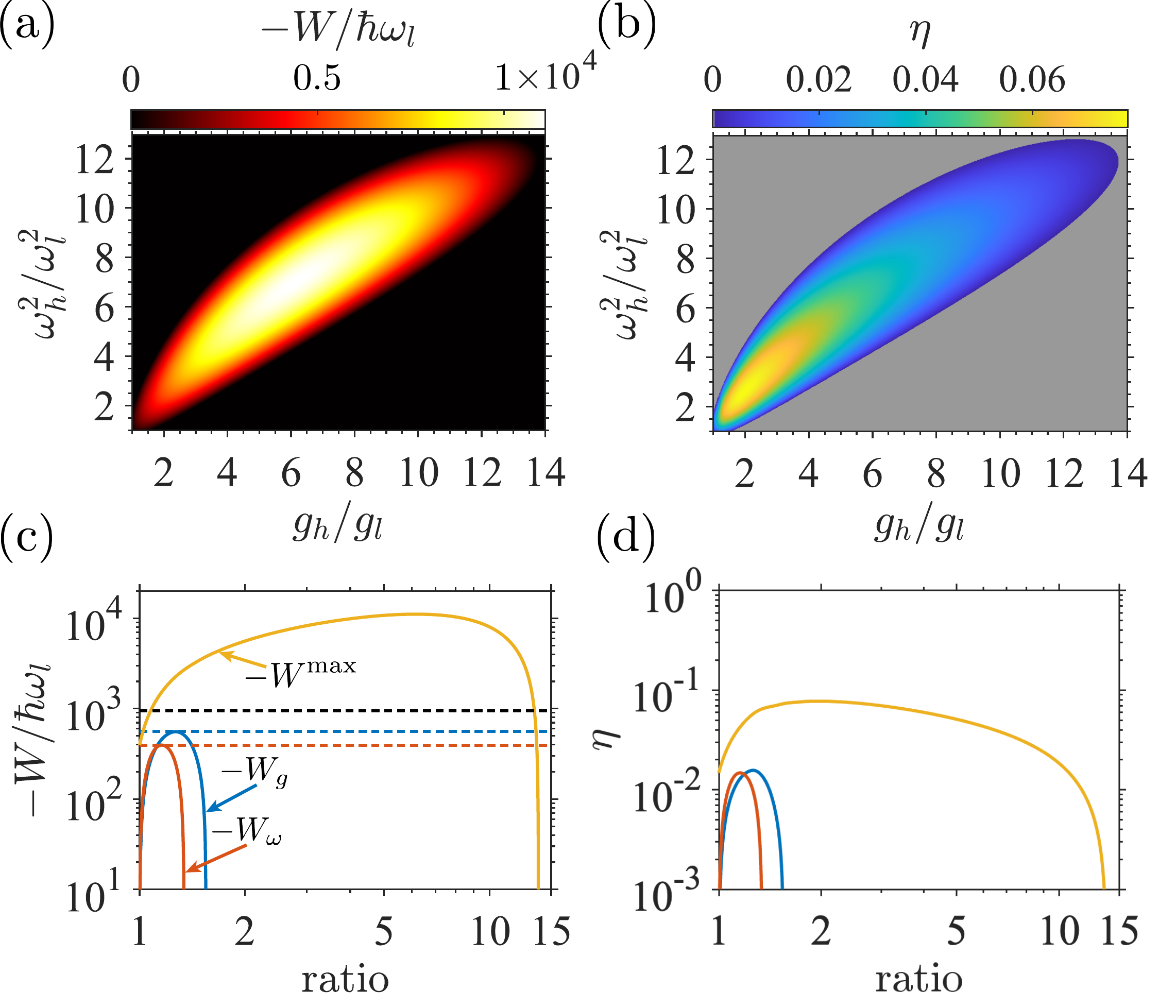}
\caption{Performance of the two-parameter sudden quench thermo-chemical quantum Otto engine cycle in the finite temperature quasicondensate regime, evaluated via numerically exact TBA methods. Panels (a) and (b) demonstrate the net work and efficiency, respectively, as a function of the ratio of both quenched parameters. The parameter values for the low energy equilibrium state are chosen to match those used in Fig.~\ref{fig:GP_T0}, but at a finite dimensionless temperature of $\tau_0\!=\!1.2\!\times\!10^{-2}$ (see text). Further, we utilize a fixed temperature ratio between the high and low energy equilibrium states of $T_h/T_l\!=\!1.33$, making this a thermo-chemical quantum Otto engine. We observe a maximum net work of $-W/\hbar \omega_l \!\simeq \!1.1\!\times\! 10^4$, corresponding to $-W/N_l\hbar \omega_l \!\simeq \!5.5$, and an efficiency at maximum net work of $\eta \!\simeq\!0.04$.
Panels (c) and (d) display a comparison between the maximum net work and efficiency at maximum net work of the two parameter Otto engine, respectively, against the single-parameter Otto engines, as previously shown for the zero temperature quasicondensate system in Fig.~\ref{fig:GP_T0_compare}.
}
\label{fig:GP_Performance_thermochem}
\end{figure}

Performance of the two-parameter thermo-chemical quantum Otto engine, in terms of net work and efficiency, for a harmonically trapped 1D Bose gas in the quasicondensate regime is shown in Fig.~\ref{fig:GP_Performance_thermochem} (a) and (b). Here, we use the same total net particle exchange of $\Delta N \!=\! 200$ as in Fig.~\ref{fig:GP_T0}, but with an additional temperature ratio between the high and low energy equilibrium states of $T_h/T_l\!=\!1.33$. We note that, in Ref.~\cite{watson2025quantum}, it was found  that---for a harmonically trapped system (as opposed to the uniform case)---it was essential to enable diffusive contact and chemical work in order that net beneficial work could be extracted from the interaction-driven Otto engine under a sudden quench.

We observe that, when compared to the pure chemical quantum Otto engine, operating at $T=0$ and shown in Fig.~\ref{fig:GP_T0}, the two-parameter thermo-chemical quantum Otto cycle investigated here operates as an engine for a broader range of both quenched parameter ratios. This difference stems from the fact that we are incorporating a finite temperature difference between the high and low energy equilibrium states, in addition to the particle number difference. This increase to temperature in the high energy equilibrium state broadens the atomic density profile, increasing the second moment of the density distribution $\langle x^2 \rangle$, and hence resulting in an enhancement to the net work extracted from the volumetric sub-cycle (see Eq.~\eqref{eq:work_bosegas}).

The interaction-driven sub-cycle is similarly altered when operating at finite temperature. However, the changes arising in this sub-cycle due to finite-temperature operation represent a smaller alteration to the net work and efficiency compared with the modifications in the volumetric sub-cycle, as finite-temperature corrections to the local second-order correlation function are minimal in the finite-temperature quasicondensate regime \cite{kerr2023analytic}.
We additionally note that, upon reducing the temperature ratio between the high and low energy equilibrium states to the point where $T_h/T_l\!=\!1$, we would effectively reproduce the results shown for the zero temperature quasicondensate system, investigated in Fig.~\ref{fig:GP_T0}, but including corrections to the correlation function and density profile arising from finite temperature effects \cite{kheruntsyan2005finite,kerr2023analytic}.

In panels (c) and (d) of Fig.~\ref{fig:GP_Performance_thermochem}, we contrast the maximum work and efficiency at maximum work of the two-parameter Otto engine against the performance of the single-parameter Otto engines in the finite-temperature quasicondensate regime. Through this, we observe a significant enhancement to the performance when utilizing a two-parameter quench, similar to what was observed in zero-temperature quasicondensate system in Fig.~\ref{fig:GP_T0_compare}. Notably, the maximum value of net work achieved from this two-parameter Otto engine cycle is again approximately an order of magnitude greater than the sum of that achieved by the single-parameter Otto engine cycles, shown as the black dashed line in Fig.~
\ref{fig:GP_Performance_thermochem}
\,(c), with a similar improvement to the efficiency at maximum  work.

Though the improved performance is guaranteed for all parameter regimes of the 1D Bose gas, it is not \emph{a priori} clear whether the enhancement achieved remains as significant outside of the weakly interacting quasicondensate regime. We therefore turn to examining engine operation, utilizing the same numerically exact TBA methods, in the strongly interacting Tonks-Girardeau regime \cite{Tonks_1936,girardeau,kerr2023analytic}.

\begin{figure}[!t]
\centering
\includegraphics[width=0.48\textwidth]{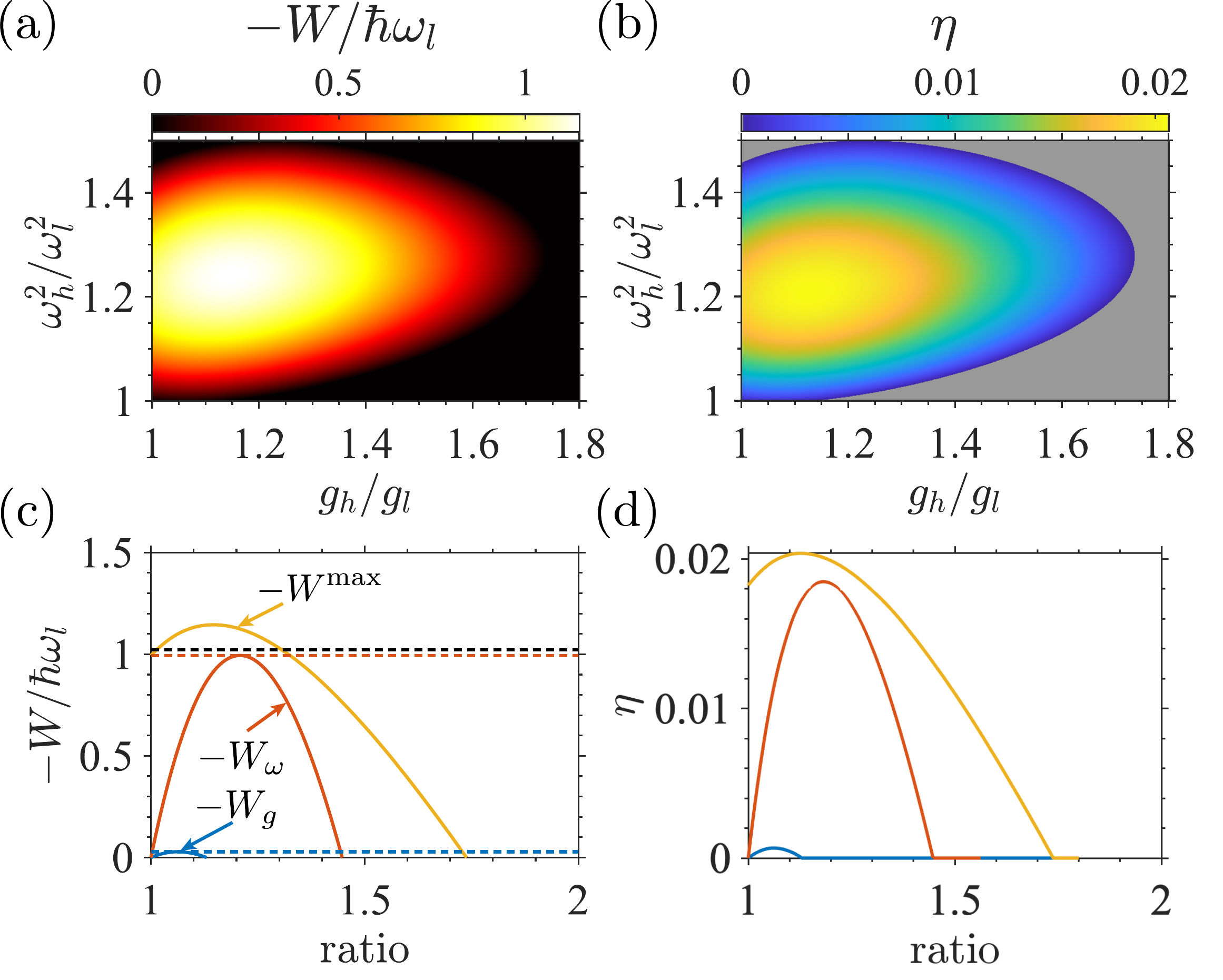}
\caption{Performance of the two-parameter sudden quench Otto engine cycle for a harmonically trapped 1D Bose gas in the strongly interacting Tonks-Girardeau regime, calculated via numerically exact TBA methods. All panels are presented in the same form as in Fig.~\ref{fig:GP_Performance_thermochem}.
The low energy equilibrium state is fixed by $N_l\!=\!20$, $\tau_0\!\simeq\!0.18$, and $\gamma_0\!\simeq\!8.5$. The high energy equilibrium state has $N_h\!=\!22$, and $T_h/T_l\!=\!2$. We observe a maximum net work of $-W/\hbar \omega_l \!\simeq \!1.1$, corresponding to $-W/N_l\hbar \omega_l \!\simeq \!5.5\times 10^{-2}$, and an efficiency at maximum net work of $\eta \!\simeq\!0.02$.}\label{fig:TG}
\end{figure}

\subsection{Tonks-Girardeau gas}
The strongly interacting ($\gamma_0 \gg 1$), low temperature ($\tau_0\!\ll\!\pi^2/(1 + 2/\gamma_0)^2$) regime \cite{kerr2023analytic}, commonly known as the Tonks-Girardeau gas \cite{Tonks_1936,girardeau}, is well approximated by a nearly ideal Fermi gas due to the strong repulsion between bosons \cite{girardeau,kheruntsyan2003pair,Sykes2008}. Experimental realization of the Tonks-Girardeau gas requires extremely strong transverse confinement, and is typically realized in the form of an array of 1D Bose gases within a 2D optical lattice at low total atom numbers \cite{kinoshita2005local,malvania2020generalized,Kinoshita1125,kinoshita2006quantum,horvath2025observingbethestringsattractive}. Yet, despite the low total number of atoms, the system remains well described by the TBA under the same local density approximation \cite{kheruntsyan2005finite,malvania2020generalized}, which we again utilize to examine operation of the two-parameter Otto engine.

Engine performance in the Tonks-Girardeau regime is demonstrated in Fig.~\ref{fig:TG}, again in terms of net work and efficiency as a function of both quenched parameter ratios, $g_h/g_l$ and $\omega_h^2/\omega_l^2$. Here, our working fluid consists of $N\!=\!20$ total atoms, at dimensionless interaction strength $\gamma_0\!\simeq\!8.5$ and dimensionless temperature $\tau_0\!=\!0.18$, meaning the system inhabits the low temperature regime of fermionization \cite{kerr2023analytic}. The Otto engine investigated here is again thermo-chemical, where $\Delta N\!=\! 2$ particles are exchanged with the reservoirs during the equilibration strokes, whereas the high and low equilibrium states are chosen to have a temperature ratio of $T_h/T_l\!=\!2$.

Notably, the enhancement to net work and efficiency achieved via the two-parameter cycle in the Tonks-Girardeau gas is not as significant when contrasted to operation in the quasicondensate regime. To compare the two-parameter engine cycle between these regimes, we consider the maximum net work normalized to the total particle number in the low energy equilibrium state, $N_l$, which varies significantly between the Tonks-Girardeau ($N_l\!=\!20$) and quasicondensate ($N_l\!=\!2000$) working fluids. Upon doing this, we find that the Otto engine operating in the Tonks-Girardeau regime achieves a maximum net work of $-W/N_l \hbar \omega_l\!\simeq\!5.5\!\times\!10^{-2}$, which is two orders of magnitude less than the same cycle operating in the quasicondensate regime, where $-W/N_l\hbar\omega_l\!\simeq\!5.5$ (see Fig.~\ref{fig:GP_Performance_thermochem} caption).

The reduction in  performance in the Tonks-Girardeau regime may be attributed to the effect of fermionization of the interaction-driven sub-cycle.
In particular, under strong interparticle interactions, the local second-order correlation function, and therefore the interaction energy of the working fluid, is reduced to near zero, dramatically reducing the net work of the interaction-driven sub-cycle. 
Further, adding particles to a fermionic system is associated with a large intake of energy, in contrast to adding particles to a bosonic system, where most particles condense into low energy states. Hence, in the Tonks-Girardeau regime, particle intake from the high energy reservoir is associated with a large energy penalty, and results in low efficiencies.

\section{Transverse-field Ising model}\label{sec:TFIM}

As a final example of this general multi-parameter enhancement, we demonstrate its applicability in the context of the  TFIM. This system consists of a 1D chain of spin-$1/2$ particles with nearest neighbor interactions in the presence of an external magnetic field. 
Unlike the harmonically trapped 1D Bose gas introduced in Sec.~\ref{sec:1D_bose_gas}, the two-parameter Otto cycle for the TFIM has no uncontrolled Hamiltonian operators $\hat{H}_0$. Instead, we have first the external magnetic field $\mathcal{h}$ acting uniformly on the spins in the $z$-direction,
\begin{equation}
    c^{(1)} \hat{\mathcal{V}}^{(1)} = - \mathcal{h} \sum_{i=1}^N \hat{\sigma}^z_i,
\end{equation}
with $\hat{\mathcal{V}}^{(1)}\equiv -\sum_{i=1}^N \hat{\sigma}^z_i$,
where $\hat{\sigma}^\alpha_i$ ($\alpha=x,y,z$) are the Pauli matrices for the $i$-th lattice site, and $c^{(1)}\!\equiv \!\mathcal{h}$.

The only other term in the Hamiltonian for this model is given by the nearest-neighbour inter-site interaction in the $x$-direction, and is represented by the two-body operator,
\begin{equation}
    c^{(2)} \hat{\mathcal{V}}^{(1)} =-J\sum_{i=1}^N \hat{\sigma}^x_i \hat{\sigma}^x_{i+1},
\end{equation}
with $\hat{\mathcal{V}}^{(2)}\equiv -\sum_{i=1}^N \hat{\sigma}^x_i \hat{\sigma}^x_{i+1}$, and $c^{(2)}\equiv J$ as the interaction strength.
Here, we choose to work in the sector with both $\mathcal{h}\!>\!0$ and $J\!>\!0$, though the analysis presented extends simply outside of this region.

\begin{figure}[!t]
\centering
\includegraphics[width=0.46\textwidth]{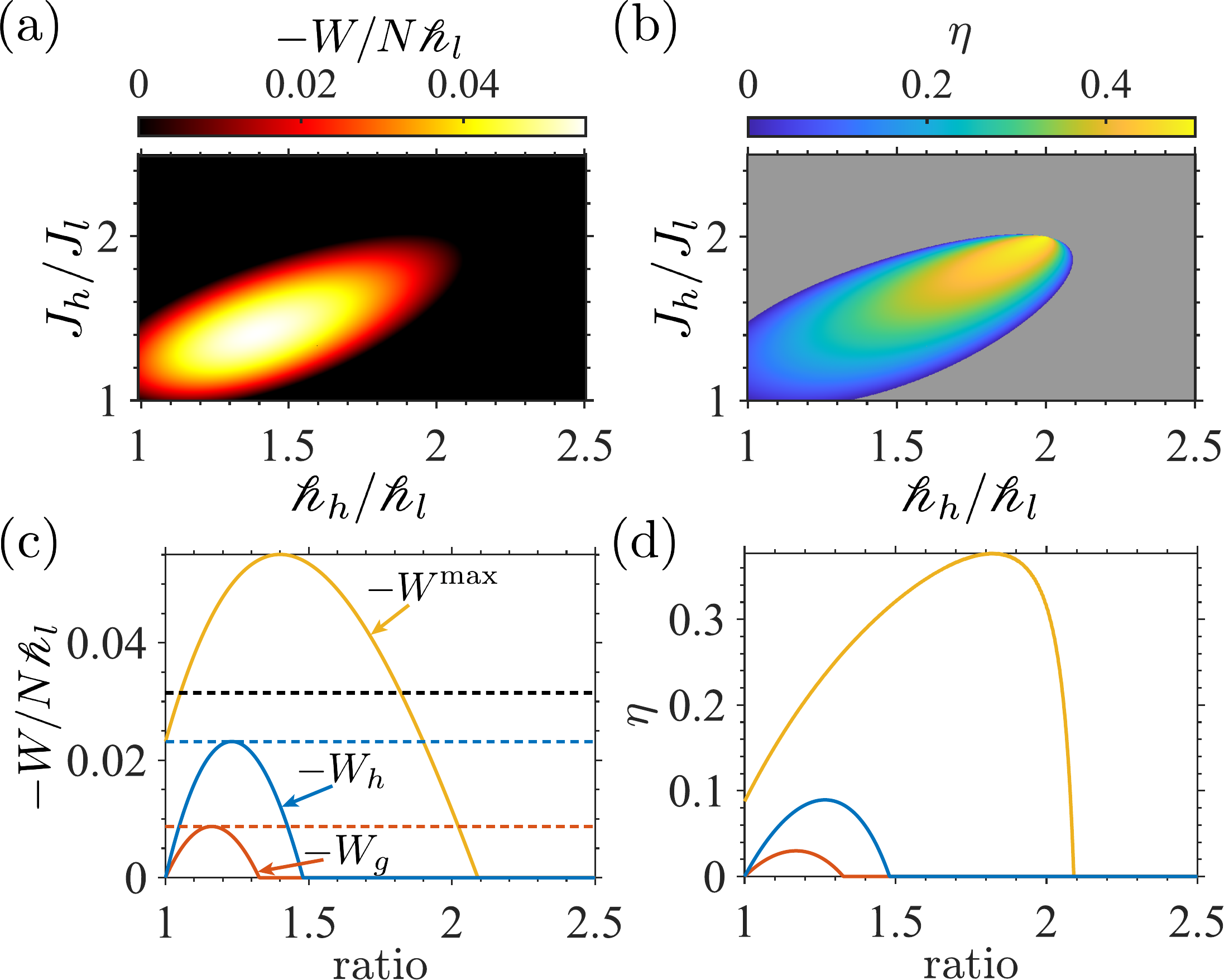}
\caption{Performance of the two-parameter sudden quench Otto engine cycle for the TFIM. All panels are presented in the same layout as in Fig.~\ref{fig:GP_Performance_thermochem}, adapted to the parameters utilized in the TFIM.
Net work is given in natural units based on the low energy magnetic field strength, $\mathcal{h}_l$, and normalized to the total particle number. The low energy equilibrium state is fixed by $J_l/\mathcal{h}_l \!=\! 1.2$, with $k_B T_l / \mathcal{h}_l \!=\!1$. The high energy equilibrium state has $T_h/T_l\!=\!2$.  We observe a maximum net work of $-W/N \mathcal{h}_l \!\simeq \!5\times 10^{-2}$, and an efficiency at maximum net work of $\eta \!\simeq\!0.3$.}\label{fig:TFIM}
\end{figure}

The two-parameter sudden quench Otto engine cycle for the TFIM corresponds to a sudden quench over both the transverse magnetic field, $c^{(1)}\!=\!\mathcal{h}$, with $\mathcal{h}_h\!>\!\mathcal{h}_l$, and the interaction strength, $c^{(2)}\!=\!J$, with $J_h\!>\!J_l$. The net work of such an engine is then given by
\begin{align}
    W &\!\simeq  (\mathcal{h}_h \!-\! \mathcal{h}_l) \left( \sum_i \langle\hat{\sigma}^z_i\rangle_h \!-\! \sum_i \langle\hat{\sigma}^z_i\rangle_l  \right)\nonumber \\
    &+\!(J_h \!-\!J_l) \left( \sum_i \langle\hat{\sigma}^x_i \hat{\sigma}^x_{i+1}\rangle_h \!-\! \sum_i \langle\hat{\sigma}^x_i \hat{\sigma}^x_{i+1}\rangle_l  \right).
    \label{eq:W_TFIM_2}
\end{align}

We focus here on the net work per particle, $W/N$, which may be calculated analytically at finite temperature in the thermodynamic limit via the Helmholtz free energy per particle, $f\!=\!F/N$.
In particular, as shown in Ref.~\cite{Pfeuty_ising}, in one dimension one may utilize the Jordan-Wigner transformation to map the spins to free fermions, which may be solved exactly to give,
\begin{equation}\label{eq:TFIM_free_energy}
   f\!=\! - \frac{1}{k_B T}\left\{\ln 2 \!+\! \frac{1}{\pi}\int_0^\pi dk \ln \left[ \cosh\left(\frac{ \Lambda_k}{k_B T}\right) \right] \right\},
\end{equation}
where $\Lambda_k =  \sqrt{J^2+\mathcal{h}^2 - 2J\mathcal{h} \cos k}$ depends on both the strength of the transverse magnetic field, $\mathcal{h}$, and on the inter-site interaction strength, $J$.

The transverse magnetization per particle, $m_z\!=\!\sum_i \langle\hat{\sigma}^z_i\rangle/N$---appearing on the first line of Eq.~\eqref{eq:W_TFIM_2}---may then be evaluated using the Hellmann-Feynman theorem \cite{hellmann1933,feynman1939forces},
\begin{equation}
     m_z = -\frac{\partial f}{\partial \mathcal{h}}= \frac{1}{\pi} \int_0^\pi dk \frac{\mathcal{h} - J \cos k}{\Lambda_k} \tanh\left( \frac{\Lambda_k}{k_B T}\right).
\end{equation}
One may utilize this method also to calculate the nearest-neighbor correlation per particle, $\sum_i \langle\hat{\sigma}^x_i \hat{\sigma}^x_{i+1}\rangle/N$, which is required for evaluating the remaining terms in Eq.~\eqref{eq:W_TFIM_2}; this takes the same form as the transverse net magnetization above, only swapping $\mathcal{h}\leftrightarrow J$ due to the symmetry of the Helmholtz free energy in these parameters.

These analytically calculable operator expectation values may then be used to evaluate the performance of the two-parameter Otto engine cycle for the TFIM in the thermodynamic limit. In Figs.~~\ref{fig:TFIM}\,(a) and (b), we show the net work per particle and the efficiency of such an engine cycle.
As we see from these figures, the net work per particle is low, being of the same order of magnitude achieved in the Tonks-Girardeau regime of the 1D Bose gas shown in Fig.~\ref{fig:TG}. The efficiency, however, is significantly higher than that achieved from a multi-parameter quench in the 1D Bose gas.

The relative increase to efficiency for the Otto engine cycle in the TFIM, when contrasted with that obtained from the 1D Bose gas, results from the fact that there is no explicit kinetic energy term present in the Hamiltonian of the TFIM. The thermal energy absorbed by the system from the hot reservoir is therefore exclusively transferred into the controllable degrees of freedom utilized in the Otto engine cycle, with no losses to the kinetic energy term, $\hat{H}_0$, which is absent in the TFIM.

Multi-parameter enhancement to both the net work and efficiency over the single-parameter engine cycles are shown in Figs.~\ref{fig:TFIM}(c) and (d), respectively. Here, as was the case for the 1D Bose gas, we observe a significant enhancement in both performance measures. In particular, we observe that the maximum net work per particle arising from the multi-parameter quench significantly out-performs the sum of that from its single-parameter counterparts.

\subsection{Small parameter quenches}

Multi-parameter enhancement to the net work for small parameter quench sizes was investigated in Sec.~\ref{sec:small_quench}. There, we found a general inequality for $-\Delta W$ in terms of the second mixed derivative of the Helmholtz free energy, given in Eq.~\eqref{eq:delta_W_smallquench}. 
The TFIM possesses a quantum critical region at finite temperature extending from the zero-temperature critical point $\mathcal{h}\!=\!J$ \cite{Sachdev_2011}, which directly affects the second mixed derivative of the Helmholtz free energy, otherwise known as the static susceptibility.
Here, we investigate multi-parameter enhancement to net work under small parameter quenches within this critical region.

We begin by defining the infinitesimal quench of the transverse field, $\mathcal{h}$, and inter-site interaction, $J$, as $d \mathcal{h} \!=\! \mathcal{h}_h \!-\! \mathcal{h}_l\!>\!0$ and $d J \!\equiv\! J_h\!-\!J_l\!>\!0$, respectively. 
Then, as we are considering operation in the thermodynamic limit, we express the condition $-\Delta W \!>\!0$, given in Eq.~\eqref{eq:delta_W_smallquench}, in terms of the net work enhancement \emph{per particle} for a two-parameter Otto engine,
\begin{equation}\label{eq:delta_W_TFIM}
     \frac{-\Delta W}{N} = 2 \,d \mathcal{h}\, d  J \frac{\partial^2 f}{\partial \mathcal{h}\partial J}>0,
\end{equation}
where $f\!=\!F/N$ is the Helmholtz free energy per particle as before. As noted in Sec.~\ref{sec:small_quench}, multi-parameter enhancement to the net work for small parameter quenches is only dependent on the properties of the high energy equilibrium state. As such, in the following analysis we omit (for clarity) the subscripts defining the high and low energy equilibrium states, instead noting here that all parameters are taken at their high energy equilibrium values.

\begin{figure}[!t]
\centering
\includegraphics[width=0.43\textwidth]{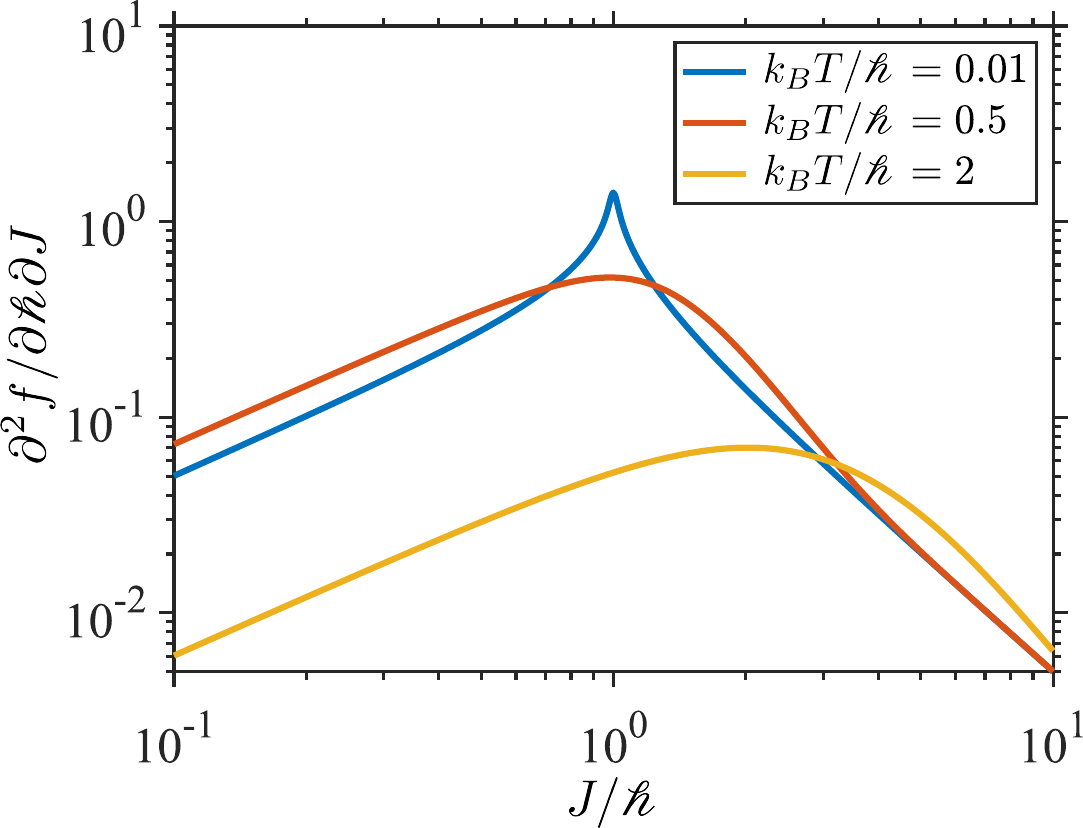}
\caption{The mixed second derivative of the free energy, otherwise known as the static susceptibility, for the TFIM. Two-parameter enhancement to the net work is directly proportional to the static susceptibility for small quenches (see text). Under decreasing temperature, a smooth peak emerges at the critical line, $J\!=\!\mathcal{h}$, indicating an increase to the two-parameter enhancement for the net work over the sum of the single-parameter sub-cycles. }\label{fig:TFIM_smallquench}
\end{figure}

Using Eq.~\eqref{eq:TFIM_free_energy} for $f$, we may evaluate its mixed second derivative,
\begin{align}\label{eq:mixed_suscept}
    \frac{\partial^2 f}{\partial \mathcal{h} \partial J} &= \frac{1}{\pi} \int_0^\pi \bigg[ \frac{\mathcal{h}J \sin^2(k)}{\Lambda_k^3}  \tanh\left( \frac{\Lambda_k}{k_B T}\right) \nonumber\\
    &-\!  \frac{(\mathcal{h}\!-\!J\cos k)(J\!-\!\mathcal{h}\cos k)}{k_B T\Lambda_k^3} \mathrm{sech}^2\left( \frac{\Lambda_k}{k_B T}\right)\bigg].
\end{align}
In Fig.~\ref{fig:TFIM_smallquench}, we show the behavior of this static susceptibility for three distinct values of the temperature. Notably, as temperature decreases, the static susceptibility shows a smooth peak at the critical line $J\!=\!\mathcal{h}$. This arises from the first term in Eq.~\eqref{eq:mixed_suscept}, which remains non-zero in the zero-temperature limit and is known to diverge at $T\!=\!0$ for $J\!=\!\mathcal{h}$. According to Eq.~\eqref{eq:delta_W_TFIM}, the presence of this peak implies an increase to the two-parameter enhancement of the net work, implying that operation of this Otto cycle is optimal at the critical line for the case of small parameter quenches.

Though the enhancement to net work decreases with increasing temperature, it remains positive at arbitrarily high temperatures. To demonstrate this, we Taylor expand the mixed second derivative, given in Eq.~\eqref{eq:mixed_suscept}, for $k_B T \!  \gg \! \mathcal{h},J$,
\begin{align}
    \frac{\partial^2 f}{\partial \mathcal{h} \partial J}\simeq \frac{4 \mathcal{h} J}{3(k_B T)^3}>0.
\end{align}
The static susceptibility is therefore strictly positive for the parameter regime we are investigating, i.e. for $\mathcal{h}\!>\!0$ and $J\!>\!0$, indicating a two-parameter enhancement to the net work at arbitrarily high temperatures.

The analysis presented in this section highlights the utility of small multi-parameter quenches for models which possess analytic expressions for their Helmholtz free energy. The simplest example of such models are free bosons or free fermions, and models which map onto these, such as the XY model, which is a generalization of the TFIM investigated here \cite{takahashi2005thermodynamics}.
Additionally, one may consider Bethe ansatz integrable models, such as the Yang-Gaudin model for spin-$1/2$ particles in 1D, or the Heisenberg spin chain \cite{takahashi2005thermodynamics}, for which there are again analytic expressions for the free energy. Notably, the Heisenberg XYZ model possesses three individually tunable parameters, and would therefore present an interesting extension to this work beyond the two-parameter case \cite{Takhtadzhan_1979,takahashi2005thermodynamics}.

\section{Conclusions}
In this work, we investigated the operation of a sudden quench Otto cycle with control over multiple external parameters. To do this, we extended the work done recently in Ref.~\cite{watson2025universal}, for sudden quenches of single parameters, to the general case of arbitrary sets of controllable parameters. Under the sudden quench approximation, the total net work separates into a sum of its constituent quenches. From this, we derived general principles dictating when the net work extracted from a two-parameter sudden quench protocol exceeds that from related single-parameter cycles taken in isolation for arbitrary quantum models, benefiting Otto engine performance. 
Further, this enhancement to the net work of the two-parameter Otto engine was found to result in an improved coefficient of performance when operating as an Otto refrigerator.

The methods introduced were applied to the case of an Otto engine cycle in an experimentally realizable harmonically trapped 1D Bose gas with contact interactions. Control over both the strength of interactions and the harmonic trapping frequency was shown to result in a region of enhanced performance when both parameters are simultaneously quenched. This was analytically illustrated in the Thomas-Fermi approximation in the simplest example of a chemical engine operating in the weakly interacting regime of the 1D Bose gas at zero temperature. 
The same enhancement to both net work and efficiency was demonstrated numerically using the thermodynamic Bethe ansatz in a finite temperature quasicondensate and the strongly interacting Tonks-Girardeau regime of the 1D Bose gas operating as a thermochemical engine.

To emphasize the universality of the methods introduced in this work to arbitrary quantum mechanical models, we applied the multi-parameter Otto cycle to the transverse-field Ising model, where we again observed a general two-parameter enhancement to both net work and efficiency.
Finally, for small quench sizes of multiple parameters, we derived a general inequality for the multi-parameter enhancement of the net work based on the Helmholtz free energy. This was then used to provide analytic insight into the operation of the two-parameter sudden quench Otto engine when applied to the transverse-field Ising model.

We emphasize that the analysis presented in this work is restricted to the sudden-quench regime, where the work depends only on the initial expectation values of the relevant operators and is therefore independent of the path taken in parameter space. For finite-time protocols, the work generally becomes path-dependent due to nonadiabatic excitations and the dynamical evolution of correlations. While the enhancement mechanism identified here is expected to persist qualitatively for sufficiently fast protocols, a quantitative description of general finite-time paths would require solving the full nonequilibrium dynamics and is beyond the scope of the present work. We also note that shortcuts to adiabaticity \cite{gueryodelin2019shortcuts} are available individually for control over both interaction strengths \cite{keller2020feshbach} and harmonic trapping frequency \cite{scale2015rhohringer,effects2020xu,Effective2021huang}, for the 1D Bose gas. It would therefore be an interesting question as to whether these shortcut methods can be combined to effectively enhance the performance of the Otto engine under a simultaneous two-parameter control.

\section*{Acknowledgments}
The authors thank Lewis Williamson and Matthew Davis for valuable discussions. This work was supported through Australian Research Council Discovery Project Grant Nos. DP190101515 and DP240101033.

\appendix


\section{Sudden quench Otto cycle}\label{app:Otto_cycle}

\begin{figure}[!t]
\centering
\includegraphics[width=0.45\textwidth]{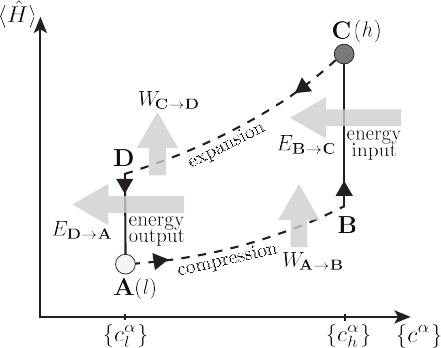}
\caption{Internal energy of the working fluid, $\langle \hat{H} \rangle$, in an interaction-driven quantum many-body Otto engine cycle operating between two interaction strengths $c_l$ and $c_h$ and in cyclic connection ($\mathbf{B}\!\to\!\mathbf{C}$ and $\mathbf{D}\!\to\!\mathbf{A}$) to two reservoirs denoted (\textit{l}) for the low energy and (\textit{h}) for the high energy state of the working fluid. Unitary work strokes $\mathbf{A}\!\to\!\mathbf{B}$ and $\mathbf{C}\!\to\!\mathbf{D}$ are denoted via dashed lines to signify the fact that these strokes are accomplished via a sudden quench rather than by passing through the intermediate equilibrium states tracing these lines.}
\label{fig:interaction_Engine_Diagram}
\end{figure}

The multi-parameter quantum Otto cycle, operating between two reservoirs denoted $(h)$ (higher energy) and $(l)$ (lower energy) in Fig.~\ref{fig:interaction_Engine_Diagram}, consists of the following four strokes,
\begin{itemize}
\vspace{-0.2cm}
\item[(1)] \textit{Unitary compression}, $\mathbf{A}\!\to\!\mathbf{B}$: the working fluid, which is initially in an equilibrium state $\hat{\rho}_l$ at a set of Hamiltonian strength parameters $\{c_l^{(\alpha)}\}$, is disconnected from the reservoir $(l)$ and has its strength parameters suddenly quenched from $\{c_l^{(\alpha)}\}\!\to\!\{c_h^{(\alpha)}\}$, with $c_h^{(\alpha)}\!>\!c_l^{(\alpha)}$ for all $\alpha$, and energy difference $\langle \hat{H}\rangle_{\mathbf{B}} \!-\!\langle\hat{H}\rangle_{\mathbf{A}}\!>\!0$. This means that the work $W_{\mathbf{A}\to\mathbf{B}}\!=\!\langle \hat{H}\rangle_{\textbf{B}} \!-\!\langle \hat{H}\rangle_{\textbf{A}}\!>\!0$ is done on the fluid. Here, $\langle \hat{H}\rangle_{\textbf{J}}$ is the expectation value of the total Hamiltonian given by Eq.~\eqref{eq:hamiltonian_multiparameter}, i.e., the total internal energy of the system, in state $\textbf{J}=\{\textbf{A,B,C,D}\}$ shown in Fig.~\ref{fig:interaction_Engine_Diagram}.
\vspace{-0.24cm}
\item[(2)] \textit{Thermalization with reservoir} (\textit{h}), $\mathbf{B}\!\to\!\mathbf{C}$: the working fluid, now in an out-of-equilibrium state, is connected to reservoir (\textit{h}) and is allowed to equilibrate while keeping the strengths, $\{c^{(\alpha)}_h\}$, constant. The working fluid takes in energy $E_{\mathbf{B}\to\mathbf{C}}\!=\!\langle \hat{H}\rangle_{\textbf{C}} \!-\!\langle \hat{H}\rangle_{\textbf{B}}\!>\!0$ from the reservoir. 
\vspace{-0.24cm}
\item[(3)] \textit{Unitary expansion,} $\mathbf{C}\!\to\!\mathbf{D}$: working fluid, now in an equilibrium state described by $\hat{\rho}_h$, is decoupled from reservoir (\textit{h}) and has its strength parameters suddenly quenched $\{c^{(\alpha)}_h\}\! \to\!\{c^{(\alpha)}_l\}$, resulting in work $W_{\mathbf{C}\to\mathbf{D}}\!=\!\langle \hat{H}\rangle_{\textbf{D}} \!-\!\langle \hat{H}\rangle_{\textbf{C}}\!<\!0$ done by the fluid.
\vspace{-0.24cm}
\item[(4)] \textit{Thermalization with reservoir} (\textit{l}), $\mathbf{D}\!\to\!\mathbf{A}$: 
the nonequilibrium working fluid is connected to reservoir (\textit{l}), allowing for energy exchange at constant $\{c^{(\alpha)}_l\}$, thus ejecting energy $E_{\mathbf{D}\to\mathbf{A}}\!=\!\langle \hat{H}\rangle_{\textbf{A}} \!-\!\langle \hat{H}\rangle_{\textbf{D}}\!<\!0$ into the reservoir, and returning to its original equilibrium state $\hat{\rho}_l$.
\end{itemize}

The net work of this sudden quench Otto engine cycle is given by Eq.~\eqref{eq:Work_generalised} in the main text.
Such a cycle generates net beneficial work (done by the fluid) when $W \!<\!0$, with a generalised engine efficiency given by Eq.~\eqref{eq:efficiency_generalised}. Notably, this generalized efficiency accounts for the fact that the energy exchange with the reservoirs may take any form (e.g. heat, chemical work, etc.).

\section{Refrigerator enhancement}\label{app:refrigerator}

\begin{figure}[!t]
\centering
\includegraphics[width=0.45\textwidth]{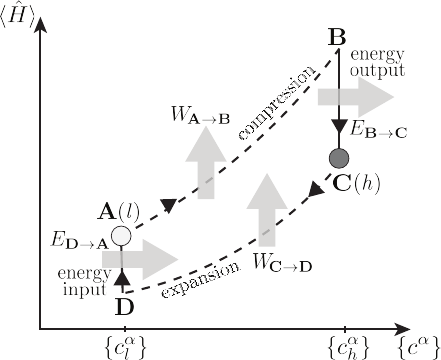}
\caption{Refrigerator operation for the two-parameter Otto cycle. For this protocol, the Otto cycle is the same as in Fig.~\ref{fig:interaction_Engine_Diagram} for the engine, except that now the energy flow between the working fluid and reservoirs is reversed \cite{SchroederD_ThermalPhysics}, meaning equilibration with the low energy reservoir takes in energy, i.e. $E_{\mathbf{D}\to\mathbf{A}}\!>\!0$, implying that the reservoir ($l$), that thermalizes the working fluid to its low-energy thermal equilibrium state, is being cooled down by the working fluid.
}
\label{fig:Refrigerator}
\end{figure}

If the inequality $-\Delta W\!>\!0$ is valid for a particular quantum model operating in an Otto cycle, it remains valid regardless of whether the Otto cycle operates as an engine, or one of the alternative protocols (e.g., refrigerator, thermal accelerator, or heater \cite{watson2025quantum}). Indeed, considering the refrigerator operation as one of the alternative protocols \cite{SchroederD_ThermalPhysics,watson2025quantum}, shown schematically in  Fig.~\ref{fig:Refrigerator}, we first note that the key parameter is the coefficient of performance ($\mathrm{CoP}$), given by
\begin{equation}\label{eq:CoP_2}
    \mathrm{CoP}_{\mathrm{two-param.}} = \frac{E_{\mathbf{D}\to\mathbf{A}}}{W}.
\end{equation}
Inspecting the combination of the two single-parameter Otto cycles, we may assign a coefficient of performance to the operation of both single-parameter cycles connected to the same high and low energy reservoirs. This combined coefficient of performance is therefore given by
\begin{equation}\label{eq:CoP_1}
    \mathrm{CoP}_{\mathrm{one-param.}} \!=\!\frac{E_{\mathbf{D}\to \mathbf{A}}^g + E_{\mathbf{D}\to \mathbf{A}}^\omega}{W^g + W^\omega}.
\end{equation}
Assuming that the net work of the two-parameter Otto cycle exceeds the sum of the single-parameter cycles, i.e. assuming $-\Delta W\!=\!-W-(-W^g+-W^h)\!>\!0$, and recalling that $W\!+\!E=0$ for cyclic operation, where $E\!=\!E_{\mathbf{B}\to\mathbf{C}}\!=\!E_{\mathbf{D}\to\mathbf{A}}$ (see Sec.~\ref{sec:mutli_param}), upon subtracting the net work of the single-parameter cycles from that of the two-parameter cycle, we find that $E\!-(\!E^g\!+\!E^\omega)>0$.

Refrigerator operation assumes that we have a finite energy intake from the high energy reservoir, as energy must be extracted for cooling to occur \cite{SchroederD_ThermalPhysics}. This means that the energy intake $E_{\mathbf{B}\to\mathbf{C}}$ must be positive. We may then subtract this positive contribution from the inequality $E\!-(\!E^g\!+\!E^\omega)>0$ to arrive at
\begin{equation}
    E_{\mathbf{D}\to\mathbf{A}}\!-\!(E_{\mathbf{D}\to\mathbf{A}}^g\!\!+\!\!E_{\mathbf{D}\to\mathbf{A}}^\omega)\!>\!0.
\end{equation}
Hence, the numerator of the two-parameter coefficient of performance, given in Eq.~\eqref{eq:CoP_2}, exceeds that of the one-parameter coefficient of performance in Eq.~\eqref{eq:CoP_1}.

Likewise, as we are assuming that the working fluid satisfies the inequality $-\Delta W\!=\!-W-(-W^g+-W^h)\!>\!0$, 
the denominator of the two-parameter coefficient of performance, Eq.~\eqref{eq:CoP_2}, is strictly less than that of the one-parameter case, Eq.~\eqref{eq:CoP_1}. Hence, we find that, for the case where net work of the two-parameter Otto refrigerator cycle exceeds that of the sum of the single-parameter refrigerator cycles,
\begin{equation}
    \mathrm{CoP}_{\mathrm{two-param.}}\!>\!\mathrm{CoP}_{\mathrm{one-param.}}.
\end{equation}
meaning that refrigerator operation, like engine operation, is enhanced for the two-parameter protocol over the combined effects of single-parameter operation.

\section{Thomas-Fermi approximation}\label{app:thomas_fermi}
Here, we derive the various analytic formulas employed in the main text for the investigation of a quantum Otto engine cycle in the quasicondensate regime. In particular, we utilize the Thomas-Fermi approximation for the density profile \cite{pitaevskii2016bose,pethick2008bose},
\begin{equation}
    \rho(x) = \rho(0) \left(1 - \frac{x^2}{R_{\mathrm{TF}}^2}\right),
\end{equation}
where $\rho(0)\!=\!(9 m N^2 \omega^2/32 g)^{1/3}$ is the peak density at the trap center (i.e. $x=0$), and $R_{\mathrm{TF}}\!=\!(3Ng/2m\omega^2)^{1/3}$ is the 1D Thomas-Fermi radius. This approximation is valid for the ground state density profile in the weakly interacting quasicondensate regime, and remains a good approximation at finite but sufficiently low temperatures, i.e. for $\tau_0\!\ll\!2\gamma_0$ with $\gamma_0\!\ll\!1$ \cite{kerr2023analytic,kheruntsyan2005finite,pitaevskii2016bose,pethick2008bose}.

The volumetric Otto engine cycle explored in the main text required evaluation of the atomic position variance, given by
\begin{align}
   \langle x^2 \rangle &= \int dx \rho(x) x^2.
\end{align}
A straightforward calculation based on the Thomas-Fermi approximation introduced above gives
\begin{equation}
    \langle x^2 \rangle =   \frac{N }{5}  R_{TF}^{2}.
\end{equation}
This formula was utilized recently to explore the volumetric Otto engine cycle under a sudden quench in Ref.~\cite{watson2025universal}.

To evaluate the performance of the interaction-driven Otto engine, we require the total correlation function, defined in the main text as
\begin{equation}
    \langle \hat{\overline{G_2}}\rangle = \int dx \,\langle \hat{\Psi}^\dagger(x) \hat{\Psi}^\dagger(x) \hat{\Psi}(x) \hat{\Psi}(x) \rangle.
\end{equation}

To progress, we introduce the normalized two-body correlation function, which is analytically tractable in the various asymptotic regimes that the 1D Bose gas possesses \cite{kheruntsyan2003pair},
\begin{align}
    g^{(2)}(x) &= \frac{  \langle \hat{\Psi}^\dagger(x) \hat{\Psi}^\dagger(x) \hat{\Psi}(x) \hat{\Psi}(x \rangle}{\rho(x)^2}.
\end{align}
Rearranging the above, we integrate to find an expression for the total integrated correlation function,
\begin{equation}
    \langle \hat{\overline{G_2}} \rangle = \int dx \, g^{(2)}(x)\rho(x)^2.
\end{equation}
We then utilize the fact that the normalized correlation function, $g^{(2)}(x)$, only slowly varies over the atomic density profile. As such, we may evaluate the total correlation as,
\begin{equation}
    \langle \hat{\overline{G_2}} \rangle \simeq g^{(2)}(0) \int dx \, \rho(x)^2,
    \label{G2_approx}
\end{equation}
which was first utilized in Ref.~\cite{kheruntsyan2003pair}, and remains a good approximation for the total correlation over the entire parameter space of the 1D Bose gas.

Next, in the weakly interacting quasicondensate regime, we know that the correlation function is well approximated by the totally coherent value of $g^{(2)}(0)\simeq 1$. Indeed, though a more detailed expression is known \cite{kheruntsyan2003pair,kerr2023analytic}, this represents a higher-order correction, and does not provide additional insights to the simple explanation given in the main text. 

Finally, we utilize the Thomas-Fermi density profile to evaluate the squared density profile in Eq.~\eqref{G2_approx} as
\begin{equation}
     \int dx \, \rho(x)^2 = b N \rho(0),
\end{equation}
where $b\equiv \,\int dx \, \rho(x)^2  /  N \rho(0)$ is a constant factor determined by the density profile, and is given by $b\!=\!4/5$ in the quasicondensate regime. We then arrive at the final expression for the total integrated correlation function utilized in the main text,
\begin{equation}
    \langle \hat{\overline{G_2}} \rangle \simeq  b N \rho(0).
\end{equation}

\end{document}